\documentclass[11pt,preprint2]{aastex}
\usepackage{latexsym}
\usepackage{graphics}

\def\hide#1{}
\def\hh{h^{-1}}

\begin{document}

\def\dim#1{\mbox{\,#1}}
\def\HI{{\rm HI}}
\def\HII{{\rm HII}}
\title{Large-Scale Simulations of Reionization}
\author{Katharina Kohler \altaffilmark{1,2,3}, Nickolay Y.\ Gnedin \altaffilmark{2,3}, and Andrew J. S. \ Hamilton\altaffilmark{1}}

\altaffiltext{1}{JILA, University of Colorado, Boulder, CO 80309, USA } 
\altaffiltext{2}{Fermilab, Batavia, IL 60510, USA; kkohler, gnedin@fnal.gov}
\altaffiltext{3}{Most of the work on this project was done while KK and NG were at CASA, University of Colorado, Boulder, 80309}

\begin{abstract}
We use cosmological simulations to explore the large-scale effects of reionization. Since reionization is a process that involves a large dynamic range - from galaxies to rare bright quasars - we need to be able to cover a significant volume of the universe in our simulation without losing the important small scale effects from galaxies. Here we have taken an approach that uses clumping factors derived from small scale simulations to approximate the radiative transfer on the sub-cell scales. Using this technique, we can cover a simulation size up to $1280 \hh\dim{Mpc}$ with $10 \hh\dim{Mpc} $ cells. This allows us to construct synthetic spectra of quasars similar to observed spectra of SDSS quasars at high redshifts and compare them to the observational data. These spectra can then be analyzed for HII region sizes, the presence of the Gunn-Peterson trough, and the Lyman-$\alpha$ forest. 

\end{abstract}


\section{Introduction}
The term ``reionization'' is used to describe the process that turned the once neutral universe into the universe we observe today, where the neutral fraction of the gas is less than $10^{-5}$. Around $z\sim1100$, the expansion of the universe cooled the Cosmic Microwave Background to the point where it could no longer keep the gas ionized, and hydrogen recombined to a neutral fraction of $\sim 0.9999$. At this point fluctuations in density, stemming from quantum fluctuations in the early universe, were imprinted on the CMB in the form of temperature fluctuations. These density variations continued to collapse after recombination and started forming the first gravitationally bound objects like galaxies and quasars. Radiation from these first objects later began to reionize the universe. A lot of work has been done on reionization starting with Giroux \& Shapiro in 1996 and continued in many papers, including but not limited to: Tegmark et al. 1997; Gnedin \& Ostriker 1997; Haiman \& Loeb 1998; Gnedin 2000; Miralda-Escud\'{e}, Haehnelt \& Rees 2000; Loeb \& Barkana 2001; Bruscoli, Ferrara \& Scannapieco 2002; Haiman 2002; Lidz et al. 2002; Hui \& Haiman 2003; Whyithe \& Loeb 2003.

Before reionization, bubbles of ionized material formed around objects emitting ionizing radiation in high density regions. These regions of ionized gas were still separated from each other by neutral high density gas, because a large number of photons are needed to keep high-density regions ionized. The abundance of ionizing photons increased with time as more objects formed, consequently creating more ionized bubbles. The result was a universe containing high density regions that were ionized close to ionizing sources and neutral high-density regions further away from sources, embedded in lower density gas that was becoming increasingly ionized.

Eventually the ionized regions expanded more and started to overlap. During the overlap period the mean free path of ionizing photons increased drastically, because more sources were visible in each ``bubble'', so that all the low and average density gas was ionized quickly. The overlap period was brief and left small, high density pockets of neutral gas. The regions around quasars however, usually assumed to be of high-density, were highly ionized. The ionizing intensity still continued to rise and the neutral regions slowly decreased. This period is called post-overlap (Miralda-Escud\'{e}, Haehnelt \& Rees 2000).

Reionization can be probed effectively with high-redshift quasar absorption spectra. In particular, a Gunn-Peterson trough (Gunn \& Peterson 1973) should appear at redshifts near and above the epoch of reionization, because of the higher neutral fraction of distributed gas. Several quasars have now been observed at high enough redshifts ($z > 6$) to probe the epoch of reionization (Fan et al. 2002; Songaila \& Cowie 2002; White et al. 2003; Songaila 2004).

The Lyman-alpha forest observed in the spectra of these quasars is remarkably
dissimilar to the classical Lyman-alpha forest at $z$ between 2 and 4, or,
for that matter, to any other type of astrophysical spectra. The Lyman-alpha
forest at $z>5$ becomes so dense that flux manages to penetrate only within
individual gaps between blends of absorption lines. The standard cosmological
theory of the Lyman-alpha forest cannot be applied to this kind of spectra,
and, and the present moment, detailed numerical simulations appear to be the
only method to create realistically looking synthetic spectra of $z>4$
forest.

To be observable, high redshift quasars must be very bright. Because such quasars can ionize an appreciable region around them, the line of sight to them may not be typical. This bias can be hard to assess from observations alone, but can be analyzed using cosmological simulations (Mesinger \& Haiman 2004; Mesinger, Haiman \& Cen 2004).

To compare the synthetic spectra to observational data we need to include high luminosity quasars in our simulation. This means that we need to create a model of a significant volume of the universe, so that enough objects with low volume density, such as high luminosity quasars, are included. Such a task is not trivial since we have to cover a large range of scale, from stars, which are thought to be vital in the process of reionization, to extremely rare quasars of luminosities in excess of $10^{12}L_{\odot}$. The simulation has to follow many matter components (e.g. dark matter, gas, stars, and quasars) and also follow the radiative transfer in order to model such a complicated process (Gnedin 2000; Ciardi, Stoehr \& White 2003; Sokasian et al. 2003; Sokasian, Abel \& Hernquist 2001; Ciardi, Ferrara \& White 2003). 

In this paper we create a physical model of reionization in large volumes. For these large-scale simulations we have to develop a method to include the small scale structure. Here this is done with a formalism of ``clumping factors'' to approximate radiative transfer at scales smaller than the resolution of our simulation. Thus, a clumping factor model allows us to drastically expand the dynamical range of our simulations and thus create a physical model of reionization, on scales unreachable by other approaches.  

First we will describe the simulations and the method used to obtain the models for the clumping factors (Section~\ref{subsec-simtheory} and Section~\ref{subsec-clump}). Then we will demonstrate how we obtained the synthetic spectra and show some applications they can be used for (Section~\ref{sec-los} and Section ~\ref{sec-quasars}). In the last parts of the paper we will show the results obtained so far and finish with a discussion of future work.

\section{Approach}

\subsection{Simulations}
\label{subsec-simtheory}
We use the ``Softened Lagrangian Hydrodynamics'' (SLH; Gnedin 1995) code as
our main simulation tool. While the detailed methods for following gas
dynamics and the dark matter dynamics are not important in this work,
since we are dealing with large, quasi-linear scales, the main advantage of
the SLH code for this project is its ability to follow time-dependent and
spatially-resolved radiative transfer using the Optically Thin Eddington
Tensor Approximation (OTVET) of Gnedin \& Abel (2001).
We have modified the code to include the clumping factors in the ionization
and thermal balance equations. 

In simulating large volumes, such as our $256 \hh\dim{Mpc}$ to 
$128 \hh\dim{Mpc}$ boxes, it is important to include various
sources of ionizing radiation correctly. In this paper we split the
total source term in the radiative transfer equation $S$ into two
components: a ``smoothly distributed'' component $S_S$ that includes sources
that cluster on scales below our resolution, and an ``isolated sources''
component $S_I$ that accounts for sources that cluster on scales
above our resolution length. For the smoothly distributed component we
assume that the source density is directly related to the matter density
$\rho$ with a power-law relation,
\begin{equation}
  S_S(\vec{x},\nu,t) = g_\nu(t) \rho^{b_S}(\vec{x},t),
\end{equation}
where $b_S$ is the ``bias factor'' and $g_\nu$ is the spectral shape of the
smoothly distributed sources (which can be time-dependent). The isolated
sources component consists of individual sources that follow the
pre-defined luminosity function and are biased with respect to the matter
density with a bias factor $b_I$.

In this paper we assume that the only type of isolated sources present in
the universe are quasars. We adopt the parameterization of the quasar
luminosity function from Schirber \& Bullock (2003) 
\begin{equation}
  \phi(L) = {\phi_*/L_* \over 
    (L/L_*)^{\gamma_B}+(L/L_*)^{\gamma_F}},
  \label{qsolumfun}
\end{equation}
where $\phi_*$ and $L_*$ are parameters, and the faint and bright
slopes of the luminosity function are $\gamma_F=1.6$ and $\gamma_B=2.6$
respectively. Following Schirber \& Bullock (2003), we parameterize time
dependence of $\phi_*$ and $L_*$ as
\begin{eqnarray}
  \phi_* & = & \epsilon^{2.72}\times 10^{2.80+0.81(z-3)} {\rm Gpc}^{-3},
  \nonumber \\
  L_* & = & \epsilon^{-1.72}\times 10^{12.1-0.81(z-3)} L_\sun.
\end{eqnarray}

The advantage of this parameterization is that it satisfies all known
observational constraints on the quasar luminosity function and the only
freedom remaining is encapsulated into the ``effective emissivity
parameter'' $\epsilon(z)$. Schirber \& Bullock (2003) argued that from the
measurements of the proximity effect and the Lyman-alpha forest flux
decrement the value of $\epsilon(z)$ at $z\sim3$ should be about 1. In order
to extrapolate $\epsilon(z)$ to higher redshifts, we assume that it
approximately follows the star formation rate from the small box
simulations described in Gnedin (2000), and we adopt the following
analytical fit to this dependence,
\begin{equation}
\epsilon(z) = C\exp\left(-((1+z)/9)^{3}\right),
\label{eq:starformation}
\end{equation}
where the constant $C$ is chosen so that $\epsilon(3)=1$.

Thus, our procedure for computing the total source function is the
following. At each time step we compute the value for the critical quasar
luminosity for which there is at most one quasar in one resolution
element. All dimmer quasars are counted as ``smoothly distributed sources'', while all brighter quasars are counted as ``isolated sources''. We
then compute the spectral shape of the ``smoothly distributed sources''
$g_\nu$ as the sum of the stellar contribution and the contribution from
the low luminosity quasars. Then individual point sources of ionizing
radiation are distributed in the computational box with the luminosity
function given by equation (\ref{qsolumfun}). The sources are distributed
randomly throughout the computational box with the probability to be
located at a given point that is proportional to the local mass density in
$b_I$ power. For the simulations described in this paper we adopt $b_S=2$
and $b_I=3$.

The isolated sources continue to emit ionizing radiation for a pre-defined
period of time, the quasar ``lifetime''. In this paper we assume that all
quasars have the same lifetime, independent of their luminosity, although
a luminosity-dependent lifetime would be straightforward to incorporate into our method.

\subsection{Clumping Factors}
\label{subsec-clump}
Presently simulations of box sizes larger than about $10\hh\dim{Mpc}$ do not have enough spatial
resolution to adequately resolve the structure in the gas down to the
smallest scale (the so-called ``filtering'' scale, Gnedin \& Hui
1998). Our simulation has a resolution of 
$2 \hh\dim{Mpc}$ for example, which does not allow us to resolve features in the Ly-$\alpha$ forest.
Thus, the evolution of gas on spatial scales below the resolution
limit must be described approximately with a phenomenological model. Such a
model is often called ``sub-cell physics''. In the case of ionization evolution
of low density gas in the IGM, a sub-cell model can be fully described by a set
of ``clumping factors''(Gnedin \& Ostriker 1997; Madau, Haardt \& Rees 1999; Ciardi et al. 2000; Miralda-Escud\'{e} et al. 2000). 

Let us consider the ionization balance equation for hydrogen:
\begin{equation}
\frac{d}{dt}{n}_{\HI}=-3Hn_{\HI}-n_{\HI}\Gamma+R(T) n_{e} n_{\HII},
\label{eq:ionbal}
\end{equation}
where $n_{\HI}$, $n_{\HII}$, and $n_e$ are number densities of neutral
hydrogen, ionized hydrogen, and free electrons respectively, $\Gamma$ is the
photoionization rate, $R(T)$ is the hydrogen recombination
coefficient, and $H$ is the Hubble parameter. Equation (\ref{eq:ionbal}) holds at every point in the
IGM. Let us now impose a finite spatial scale on the true distribution of
cosmic gas - in our case, the finite resolution scale of a simulation will be
such a scale. Averaging equation (\ref{eq:ionbal}) over such a scale (let
us call it a ``cell''), we obtain:
$$
\frac{d}{dt}\langle{n}_{\HI}\rangle=-3H\langle n_{\HI}\rangle-\langle n_{\HI}\Gamma\rangle
$$
\begin{equation}
\phantom{AAAAAAAAAA}+\langle R(T)n_{e}n_{\HII}\rangle
\label{eq:averagedoriginal}
\end{equation}

Since numerical simulation can only deal with quantities defined within one
cell, we must express the right hand side of equation
(\ref{eq:averagedoriginal}) as a function of physical quantities averaged
over one cell, namely
\begin{equation}
\frac{d}{dt}{\tilde{n}_{\HI}}=-3H\tilde{n}_{\HI}-{C}_{I}\tilde{n}_{\HI}\tilde{\Gamma}+{C}_{R}\tilde{R}\tilde{n}_{e}\tilde{n}_{\HII},
\label{eq:cellionbal}
\end{equation}
where we use the notation such that for any physical quantity $f$ the tilde
symbol represents the average over one cell,
$$
\tilde{f} \equiv \langle f\rangle_{\rm cell},
$$
and $C_{I}$ and $C_R$ are ``clumping factors'' defined as
\begin{equation}
C_{I}=\frac{\langle n_{\HI}\Gamma\rangle}{\langle n_{\HI}\rangle\langle \Gamma\rangle},
\end{equation}
\begin{equation}
C_{R}=\frac{\langle R(T) n_{e} n_{\HII} \rangle}{\langle R(T) \rangle \langle n_{e}\rangle \langle n_{\HII}\rangle}.
\end{equation}

Analogously, the radiative transfer equation for the spatially variable ionization intensity $\tilde{J}_{\nu}$ averaged over one cell can be written as
$$
\frac{\partial \tilde{J}_{\nu}}{\partial t}+\frac{c}{a}\vec{n}\frac{\partial \tilde{J}_{\nu}}{\partial \vec{x}}+H(\nu\frac{\partial \tilde{J}_{\nu}}{\partial\nu}-3\tilde{J}_{\nu})=
$$
\begin{equation}
\phantom{AAAAAAAAA}-C_{I}\tilde{k}_{\nu}\tilde{J}_{\nu}+\tilde{S}_{\nu},
\end{equation}
where $\tilde{k}_{\nu}$ is the absorption coefficient and $\tilde{S}_{\nu}$ is the source function given by
\begin{equation}
\tilde{S}_{\nu}=\tilde{S}_{\nu,I}+\tilde{S}_{\nu,S}.
\label{equ:sourcefunction}
\end{equation}
Here $\tilde{S}_{\nu,I}$ is the source function component from individual resolved quasars and $\tilde{S}_{\nu,S}$ is the component from the smooth background.

In particular, it is worth noting that we have two clumping factors, one
in the recombination term, $C_{R}$, and another one in the ionization term,
$C_{I}$. Both clumping factors are necessary to properly account for
the evolution in ionization state of cell-averaged hydrogen number
density. 

Clumping factors cannot be derived from the large-scale simulations that
only resolve structures on scales above one cell. Thus, additional
information must be used to specify the clumping factors and close equation
(\ref{eq:cellionbal}). 

To determine the clumping factors, we use a simulation of reionization
with the small size of the computational box ($4\hh\dim{Mpc}$) with the
Softened Lagrangian Hydrodynamics (SLH) code (Gnedin 1995, Gnedin
2000). While this simulation has too small a box size to be useful for
modeling effects of bright rare quasars, it has enough spatial resolution
to follow the structure of the IGM down to the filtering scale, and thus,
can be used for computing the clumping factors on scales of
interest. Specifically, we split the 
$4\hh\dim{Mpc}$ box into 8 cubes $2\hh\dim{Mpc}$ on a side and averaged
$n_{\HI}$, $n_{\HII}$, $n_e$, $R(T)$, $\Gamma$, and their appropriate
products over the volume of each of 8 cubes. We have done this for a range 
of redshifts from $z\sim5$ to $z\sim12$, obtaining 8 data points for each 
of the two clumping factors for each redshift value. Each cube gives us a 
different value, depending on the ionization and the density in its volume. This 
method yields clumping factors that are not dependent on redshift, but only on properties 
of the gas. We then fitted both clumping factors as functions of two variables, 
the neutral hydrogen fraction $x=\tilde{n}_{\HI}/\tilde{n}_{\rm  H}$, and 
the gas density $\rho$.

Formally, averaging in equation (\ref{eq:averagedoriginal}) is performed
over the volume of one cell. However, different weightings can be used in
making the averages. For example, if we take a simple volume-weighted
averages in equation (\ref{eq:averagedoriginal}), we will be including
ionizations and recombinations not only in the low density IGM, but also in
the high density regions within the virialized halos. Thus, both clumping
factors will be large, but a significant fraction of ionizing photons will
be absorbed locally, within the parent halo of a ionizing source, so the
ionization and recombination terms in equation (\ref{eq:cellionbal}) will nearly cancel each
other. This behavior results in numerical loss of precision, and is not
desirable. 

Alternatively, we can weight the low density gas more than the high density
regions, reducing both clumping factors, but still keeping them consistent
with each other. In that case we would exclude some of the ionizations and
recombinations that take place in high density regions from equation
(\ref{eq:cellionbal}). Such an exclusion is equivalent to counting only a
fraction of ionizing photons in the ionization balance equation, and
assuming that the rest of ionizing photons are absorbed locally, within the
immediate vicinity of the ionizing source. This fraction is commonly known
as the ``escape fraction'', although the specific mathematical definition
of the escape fraction depends on the specific prescription of how averaging
is done in equation (\ref{eq:averagedoriginal}).

In this paper, we consider three different cases for performing averaging
in equation (\ref{eq:averagedoriginal}). If the clumping factors are
computed self-consistently and the
escape fraction is chosen appropriately, then the results of a simulation
should be independent of the way averaging is performed. Thus, comparing
results of simulations with different types of averaging used allows us to
estimate the uncertainties due to our final spatial resolution and
inaccuracies in parameterizing clumping factors.

\begin{figure}[h]
\begin{center}
\plotone{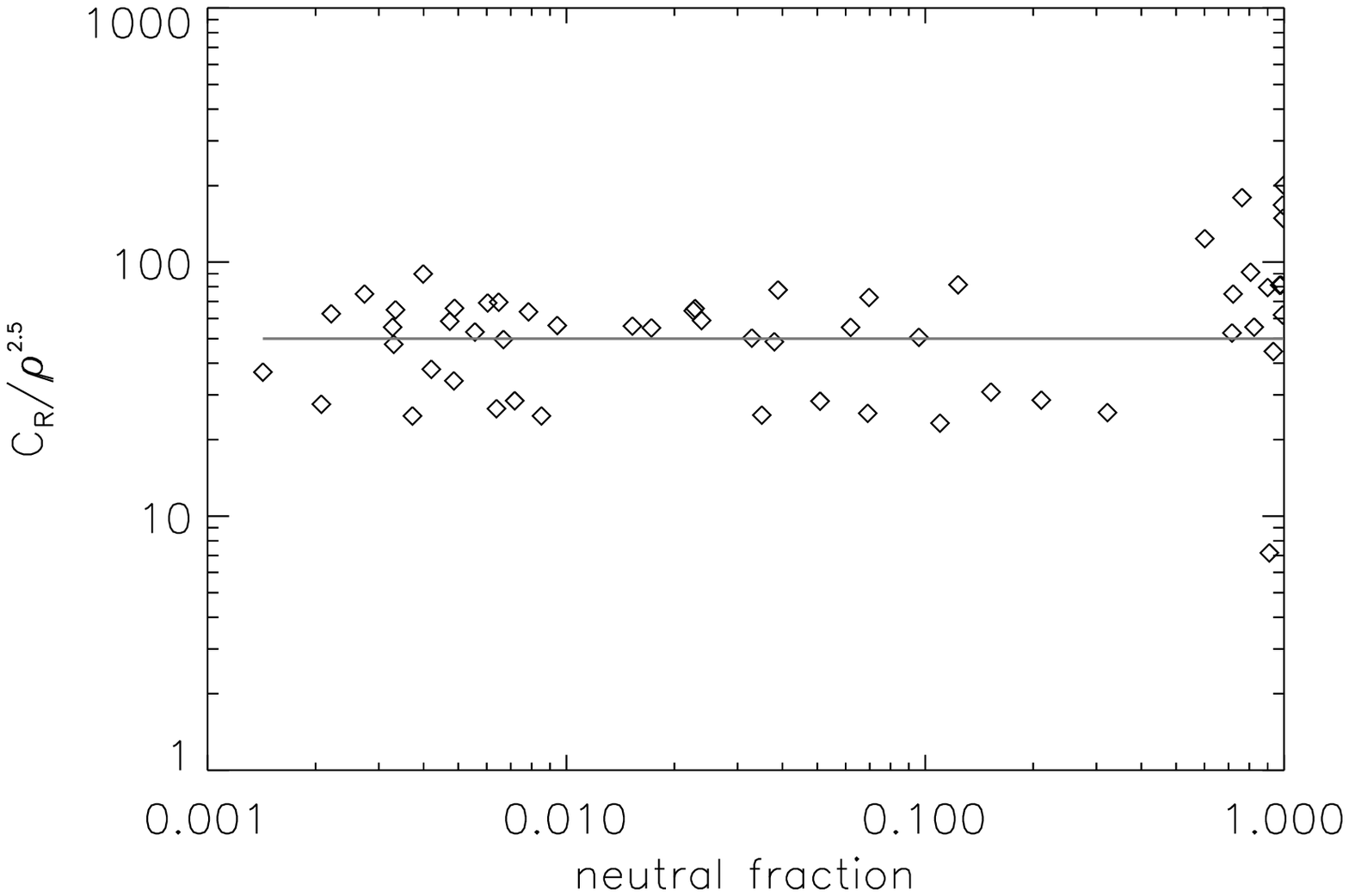}
\plotone{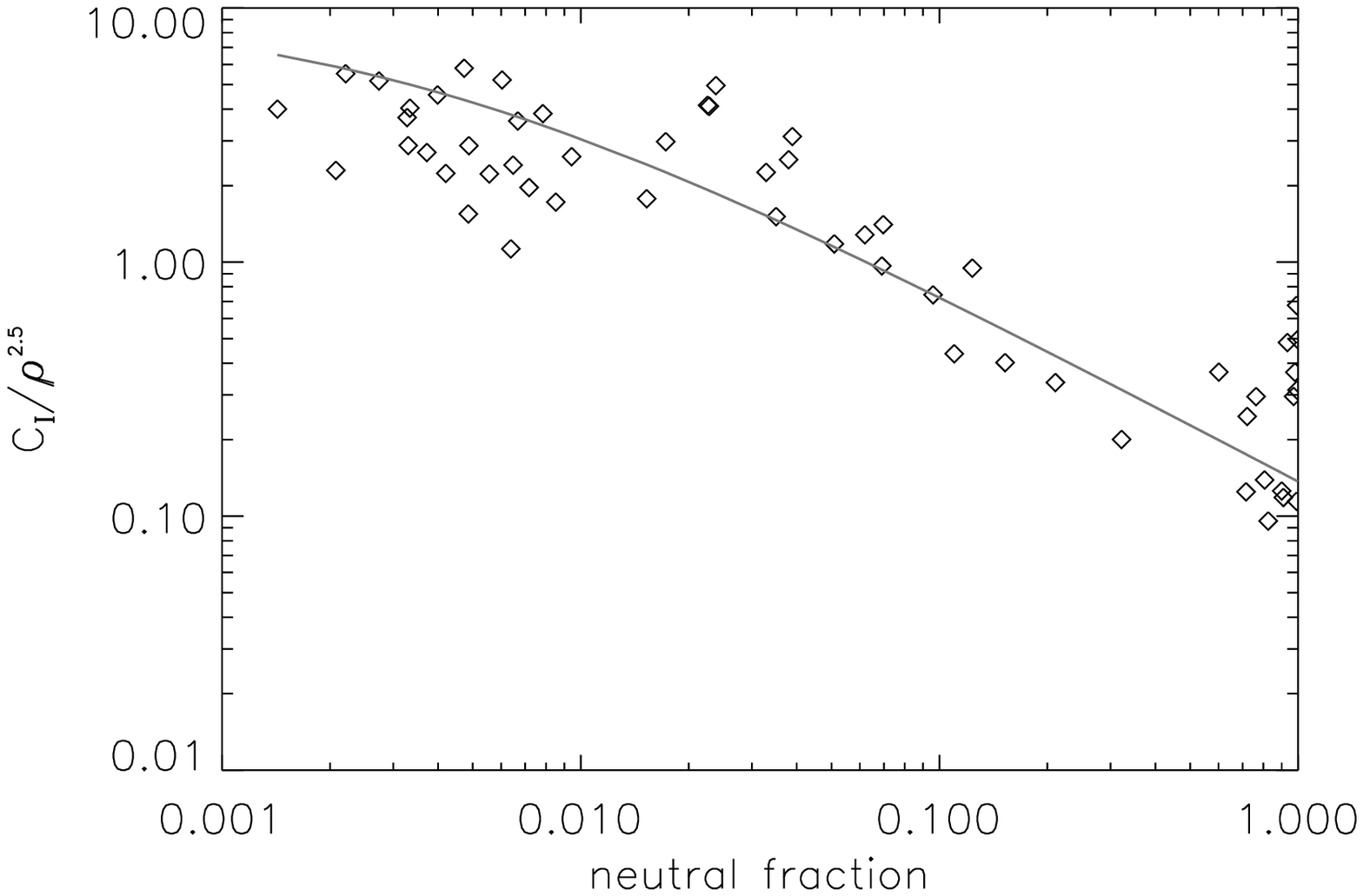}
\end{center}
\caption{Unweighted Case (Case A): $C_{R}/\rho^{2.5}$ versus neutral fraction (\textit{top}) and $C_{I}/\rho^{2.5}$ versus neutral fraction (\textit{bottom}). Functional fit to the data is shown in \textit{gray}.}
\label{fig:casea}
\end{figure}

The first case (case A) we consider is where the averaging in equation
(\ref{eq:averagedoriginal}) is volume weighted, i.e. all local absorption
counts, including the photons that are absorbed close to the source. In
this case all ionizations and recombinations are counted equally, no matter
how high the density in a cell is. The clumping factors for case A are
large, since all photons are taken into account, and the high density regions are included. The relations between the clumping factors for Case A can be seen in Figure~\ref{fig:casea}. The recombination clumping factor is not a strong function of neutral fraction, because it is dominated by high density regions. It is well approximated by a power law in density, $C_{R}\propto\rho^{2.5}$. The same power law holds for the photoionization clumping factor, but here we additionally have a dependence on neutral fraction. The photoionization clumping factor decreases with increasing neutral fraction, showing that the photoionization rate and neutral fraction are anti-correlated. 

The next two cases reduce the importance of the high density and so exclude some of the ionizations and recombinations in high density regions.

\begin{figure}[h]
\begin{center}
\epsscale{1.00}
\plotone{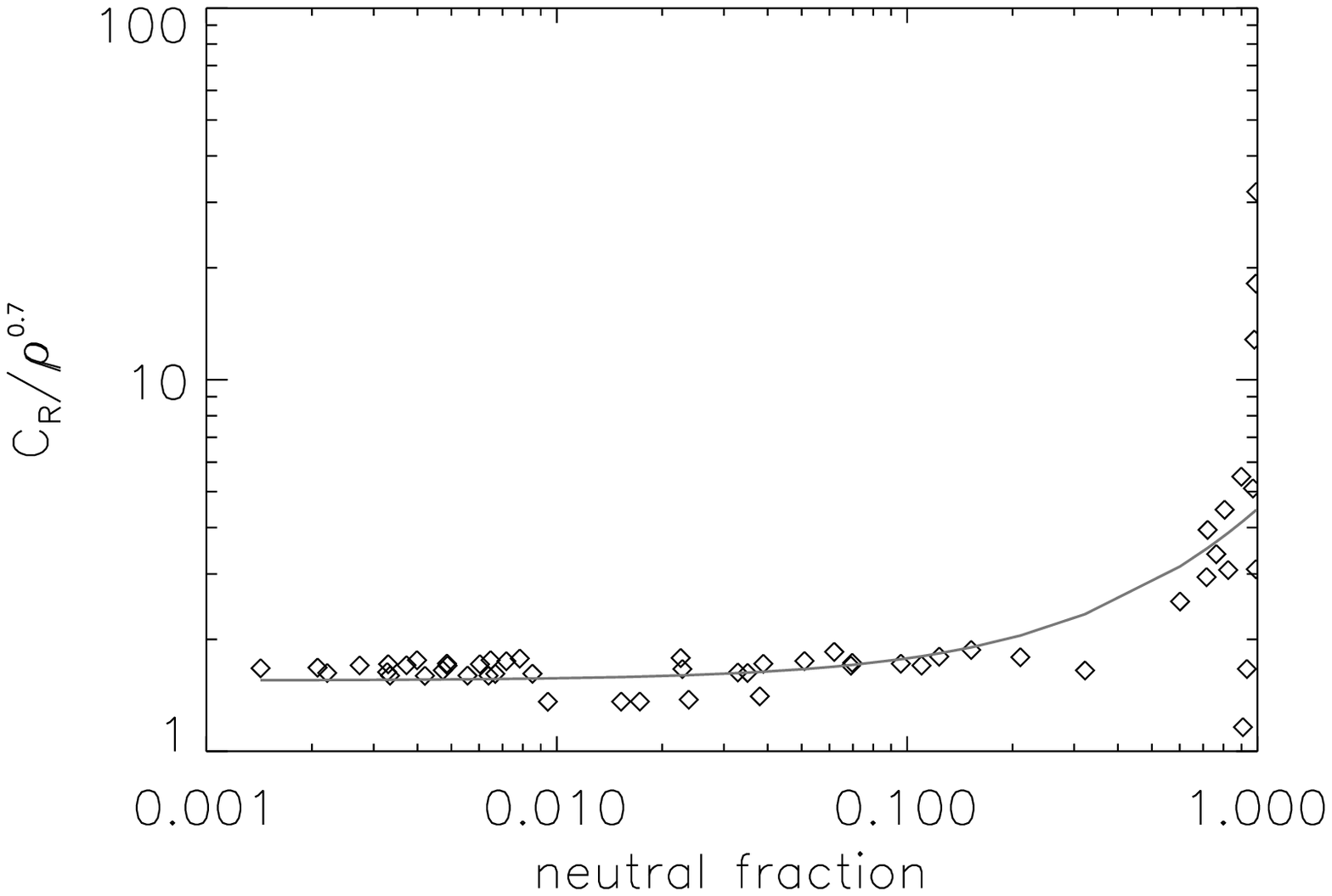}
\plotone{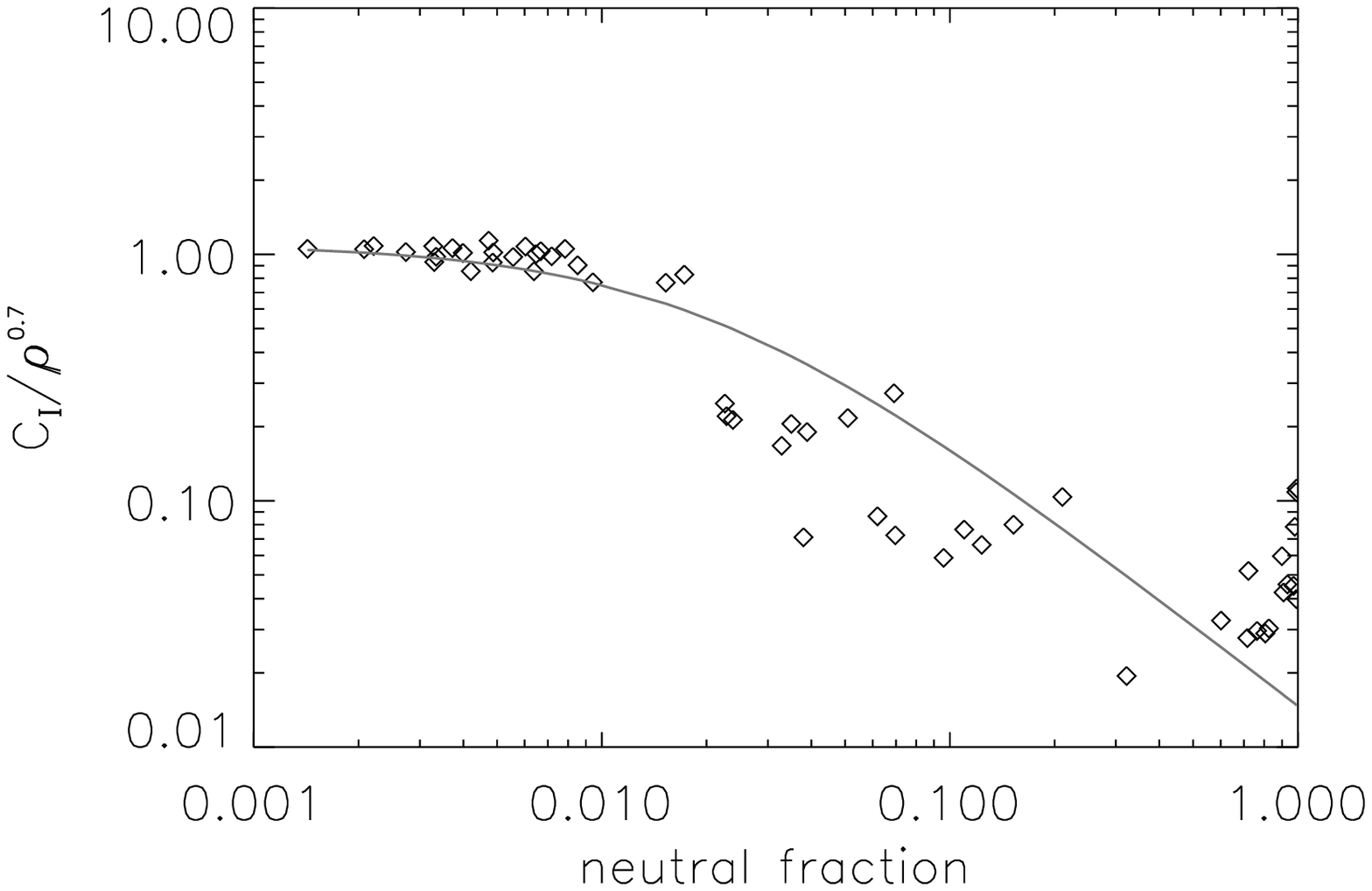}
\end{center}
\caption{Case B: $C_{R}/\rho^{0.7}$ versus neutral fraction (\textit{top}) and $C_{I}/\rho^{0.7}$ versus neutral fraction (\textit{bottom}). Functional fit to the data is shown in \textit{gray}. }
\label{fig:caseb}
\end{figure}

The second model (Case B) weights the volume by the inverse density, therefore removing some local absorption. This increases the relative weight of the low density regions, so that high density regions do not contribute as much to the volume considered for the clumping factors. The clumping factors are much smaller in this case, eliminating a lot of the numerical problems that arise in case A. Figure~\ref{fig:caseb} shows that they are of order unity, compared to order of 10 for case A. Additionally the exponent in the power law dependence of the clumping factors on $\rho$ is lower compared to case A. Both factors depend on the neutral fraction in a more complicated way. The scatter is relatively small for the recombination clumping factor and in the low neutral fraction regime of the photoionization clumping factor. Figure~\ref{fig:caseb} shows that $C_{R}$ increases with increasing neutral fraction, meaning that the regions with lower temperature and therefore higher recombination rate are more neutral, as is expected. This effect only appears when the neutral fraction is low enough, corresponding to time outputs before reionization.  $C_{R}$ does not depend on the neutral fraction below $x_{HI}\sim 0.1$, because the high density regions dominate again at high ionization fractions.

Similarly to case A, $C_{I}$ shows a decrease toward higher neutral fraction, but shows no dependence on $x_{HII}$ at low neutral fraction. This change in the slope of the data implies that the anti-correlation of the photoionization and the neutral fraction disappears when the HII regions are overlapping.

\begin{figure}[h]
\begin{center}
\epsscale{1.00}
\plotone{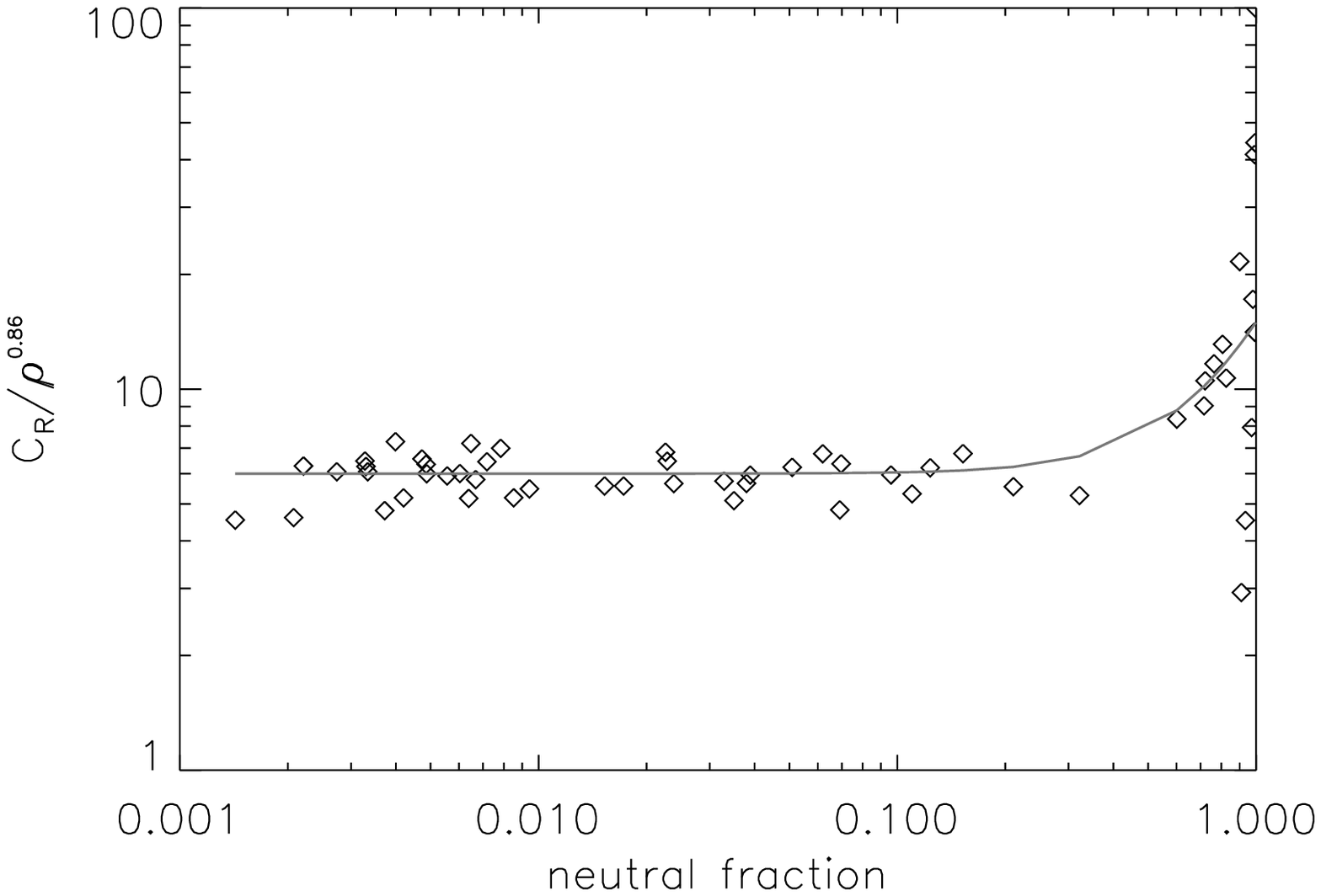}
\plotone{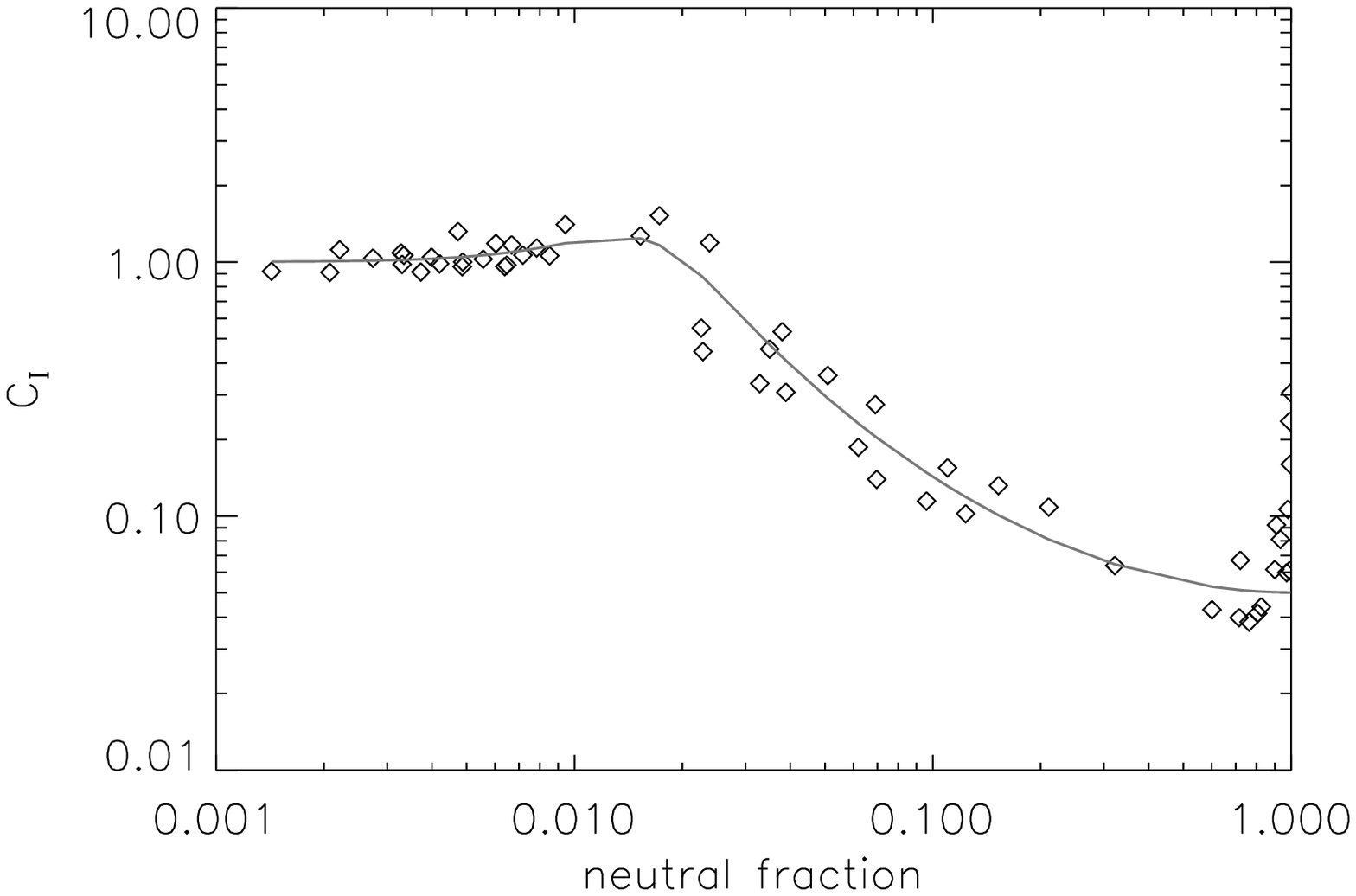}
\end{center}
\caption{Case C: $C_{R}/\rho^{0.86}$ versus neutral fraction ({\it top}) and $C_{I}$ versus neutral fraction ({\it bottom}). Functional fit to the data shown in gray}
\label{fig:casec}
\end{figure}

In the third model (Case C) all volume elements with gas density higher than the density at the virial radius were removed from the calculations of the clumping factors. Consequently, in the modeling of case C only photons that escape the high density halos are used to reionize the IGM. This removes all local ionization and absorption from the simulation. Similarly to case B the clumping factors are small compared to case A. Also, most features in Figure~\ref{fig:casec} are similar to case B in the distribution of data for case C. In particular the recombination clumping factor distribution has the same shape as $C_{R}$ for case B, even though its amplitude lies at about 6 instead of 2 in case B. 

On the other hand the photoionization clumping factor is best fit by a differently shaped function that remains nearly constant at low neutral fraction. The data for case C at higher neutral fraction closely resembles case B. These distributions shows the similarity of the behavior of their clumping factors with respect to changes in neutral fraction and density for the two density weighted models. As mentioned above all three models should be consistent with each other when the appropriate effective escape fractions are set, but the data shows that model B and model C are also similar in the details of their approximations. 

Case B appears to be the model most closely resembling the true physical situation, because it does not completely ignore the photons coming from high density regions, and also considers that many of them should get absorbed on their way out. 
There is significant scatter of the data around the fit, because clumping factors cannot be represented as simple functions of gas density and ionization state, but also depend on other properties like a distributions of sources, ionization history, clustering of sources, etc. We cannot represent this scatter of the data by simple averages of density and neutral fraction, since it is caused by variations on smaller scales. In the future, statistical treatment of the data could help to improve the fitting and take into account the stochasticity in the local properties of the IGM, but this is outside the scope of this paper.

To sum up, using these models of ``sub-cell physics'' allow us to increase the cell size of the simulation from $\dim{kpc}$ to $\dim{Mpc}$ scales while approximately retaining the effect of physical processes at small scales.

\subsection{Escape fraction}
\label{subsec-fesc}

Generally the phrase ``escape fraction'' is understood as the fraction of ionizing photons that leave a high density region surrounding a source without being absorbed locally. We define an effective escape fraction $f_{esc}$ that measures the amount of ionizing photons emitted from homogeneously distributed sources (see equation~\ref{equ:sourcefunction}). Dropping the subscript $\nu$ the smooth component of the source function for each frequency can be described with:
\begin{equation}
\tilde{S}_{S}=f_{esc}\tilde{S}^{(0)}_{S,\star}+\tilde{S}_{S,U},
\label{eqn:escapefrac}
\end{equation}
with
\begin{equation}
\tilde{S}^{(0)}_{S,\star}\sim\dot{\rho}_{\star}
\end{equation}
and
\begin{equation}
\dot{\rho}_{\star}=0.1\epsilon(z)\frac{\dim{M}_{\sun}}{\dim{yr}\dim{Mpc}^{3}}\rho_{gas},
\end{equation}
where $\tilde{S}^{(0)}_{S,\star}$ is the contribution from stars in galaxies, $\tilde{S}_{S,U}$ is the contribution to the source function from unresolved quasars, and $f_{esc}$ is the effective escape fraction. Also, $\dot{\rho}_{\star}$ is the star formation rate described by the gas density $\rho_{gas}$ and the efficiency $\epsilon$. $\tilde{S}_{S,U}$ is usually negligible compared to the stellar component, but has to be included for completeness. Both components are given by the star formation rate and thus are the same in all simulations, making $f_{esc}$ the only free parameter. 

 Equation~\ref{eqn:escapefrac} shows that any change in $f_{esc}$ will change the amount of photons present outside the high density regions and thus the average ionizing flux transmitted through the box. This in turn changes the amount of neutral hydrogen present. Thus, decreasing the escape fraction results in a lower number of photons available for ionizing the IGM and influences the time of reionization. In our simulations, we treat the escape fraction as a phenomenological free parameter, which we adjust to fit the mean transmitted flux observed in the spectra of SDSS quasars (See Section~\ref{sec-results}). 

For each of the three cases of clumping factors mentioned above, we have to adjust the effective escape fraction, since each case treats the photons emitted and absorbed in high density regions differently. First we have Case A, where we include all the local absorption in calculating the clumping factors, and secondly case B, where we weight the volume by the inverse density, increasing the influence of the lower density regions. Thus, both cases include the high density regions and the photons absorbed there. On the other hand case C does not include the high density regions and therefore less photons are used in the reionization process.

In general when escape fractions are calculated, only photons that actually escape from the source are counted. Considering that these sources are commonly positioned in high density regions in the IGM , this more general definition of escape fraction best corresponds to case C.   

Since the effective escape fraction describes the amount of photons that escape compared to the amount of photons produced, $f_{esc}$ should be unity for case A, if the star formation rate used for the simulation is correct, since all photons are counted. Our model given by equation~\ref{eq:starformation}, however, does not calculate the star formation rate perfectly realistically. It is expected that the effective escape fraction will be larger if the star formation rate is underestimated to correct for the lack of photons.
Additionally, errors in the calculation of $C_{I}$ and $C_{R}$ due to scatter in the simulated data (see Section~\ref{subsec-clump}) change the value that is required for $f_{esc}$ to yield data in agreement with observations. This problem is compounded by the limited resolution of the small-box simulation used to develop the fit for the clumping factors.

\hide{To fit the averaged flux from the simulation to the SDSS data, we produce synthesized spectra that sample the simulation box along random lines of sight. The method of obtaining these spectra is the same as the one described below for the spectra used for analysis (see Section ~\ref{sec-los}). For our analysis of \HII\ region sizes we needed to start the spectra at the position of bright quasars. However, for the computing the mean transmitted flux, we do not need to start our lines of sight at a quasar. So, instead of setting the starting position at the brightest quasars in the simulation box we could simply cast a ray from a point chosen randomly in the box at a random direction. Since it simplified the process of analysis greatly, we used the same spectra as for the \HII\ region and statistical analysis of the spectra (Section ~\ref{subsec-HIIsize} and Section ~\ref{subsec-troughs}). Thus the very high redshift ends of the lines of sight used for fitting to the SDSS data have more flux than expected, due to the flux transmitted through the \HII\ region of the quasar.}

\section{Spectra}
\subsection{Lines of Sight}
\label{sec-los}
The main purpose of our large scale simulations is to produce synthetic absorption spectra, which can be compared directly to actual spectra. First we determine the positions of the brightest quasars present in the simulation box at the chosen redshift to get starting points for the lines of sight. Beginning the spectrum at precisely the source's position is important, because only then does the \HII\ region surrounding it appear in the spectrum. This will allow us to compare our synthetic spectra to absorption spectra of high-redshift quasars where the source is used as the background lighting and derive properties of the \HII\ regions surrounding the bright quasars. The spectrum will not display the increase in transmitted flux expected for an \HII\ region at the high-z end of the spectrum if it starts at a different position.

After determining the starting point, we cast a ray from the quasar position through the box by setting a random direction and following it on a straight line through the box. To get additional resolution and a smoother spectrum, we cast the ray with a step size one quarter of the cell size. The values of density, neutral fraction and temperature are output at each step by using a weighted average of their values at the 8 closest grid-points. Because many spectra sampling different directions are necessary for statistical analysis of the HII regions and IGM properties, we created three to five spectra for each of the $50$ brightest objects in the simulation box.

Because of the large distances along the ray, the universe expands measurably while an imaginary photon of our ray of light crosses the computational box. Thus, we have to take into account the expansion of the universe along the spectrum. Accordingly, we not only have to cast the line of sight in space, but also have to consider the time evolution, using consecutive output files of the simulation. 

The spectra so obtained represent random samples of the universe between $z=6.5$, corresponding to $9120\AA$ for Lyman-$\alpha$ emission, and $z=4.0$. The starting points for the different sets of spectra were picked to be the redshifts of the most distant quasars observed and also included some lower redshifts to determine the evolution of \HII\ region sizes (starting points between $z=6.5$ and $z=5.9$). The comoving distance between $z=6.5$ and $z=4.0$ is $891 \hh \dim{Mpc}$, so depending on the direction of the line of sight and the position of the quasar a spectrum can cover a distance too long for one box size. In this case the spectrum wraps around the periodic box to cover the wavelength range from before to after reionization. The direction of the ray is chosen randomly to get around the problem of the ray passing repeatedly through exactly the same region of the simulation. This stacking of boxes is similar to the method used by Mesinger et al. (2004) and Cen et al. (1994).

\subsection{Small Scale Structure}
\begin{figure}[h]
\begin{center}
\epsscale{1}
\plotone{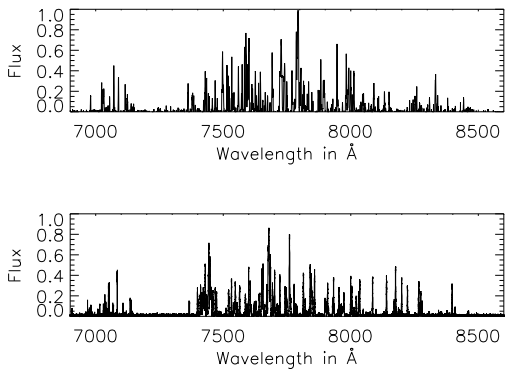}
\end{center}
\caption{Spectrum SDSSJ1030+0524 (\textit{top}) and a synthetic spectrum including noise and resolution (\textit{bottom}). Absorption by the complete Lyman series was included.}
\label{fig:spectraFAN02496}
\end{figure}

From the simulation output alone we can only extract information about large scale variations in $\rho$, $n_{HI}$, $n_{HII}$ and $T$, because the resolution of the simulation is limited. High-resolution spectra from instruments such as Keck, can reveal information about more neutral gas in between the quasar and the observer on small scales using the Lyman-$\alpha$ forest lines. For example the spectrum (Figure~\ref{fig:spectraFAN02496}) of quasar SDSS J1030 at $z=6.28$ has a resolution of $\delta\lambda = 1.63 \dim{\AA}$ whereas our resolution from the simulation is about $\delta\lambda = 9 \dim{\AA} $. 

Typically, normal sized galaxies or Lyman limit systems show Lyman-$\alpha$ features on a few $\dim{\AA}$ scale, smaller than the resolution of the simulation. To allow comparison to real observations we add, a posteriori, small scale structure due to density fluctuations below the resolution limit.

In order to generate the small scale structure along the simulated line of
sight, we compute the 1D matter power spectrum as
\begin{equation}
  P_{1D}(k) = \pi\int_k^\infty q P_{3D}(q)dq,
\end{equation}
where $P_{1D}(k)$ and $P_{3D}(q)$ are one-dimensional and three-dimensional
power spectra respectively. A similar equation can be written for the velocity fluctuations, which are needed to take into account the effects of Doppler shifts from peculiar motion. 
To get a resolution in wavelength space fine enough to be able to compare the flux to observational spectra the number of steps $n$ were chosen to be $2^{15}$. We then apply a Fast Fourier Transform to generate a real-space representation of linear overdensity $\delta_{L}$ and velocity $v_{L}$ along a line of sight on a uniform spatial mesh
with mesh spacing $\Delta x=10\hh\dim{kpc}$ (small enough so that thermal
broadening of the synthetic spectra will alleviate the dependence on this parameter).

We need to modulate the small scale structure using the large scale variations directly obtained from the simulation in density $\rho_{sim}$, temperature $T_{sim}$,
and neutral hydrogen density $n_{HI,sim}$ along the simulated line of sight to obtain the small-scale distributions.  This combines the data obtained from the simulation with a resolution on scales of a few $\hh\dim{Mpc}$ with the fluctuations implied by the power spectrum on much smaller scales.

We transform the small scale linear fluctuations $\delta_{L}$ along a line of sight using the log-normal model (Shandarin et al. 1995; Bi \& Davidsen 1997) to get 
\begin{equation}
\rho_{los}(l)=\rho_{sim}e^{\delta_{L}-\frac{\sigma^2}{2}},
\end{equation}

and then use this $\rho_{los}(l)$ to obtain the distributions for temperature and neutral hydrogen density:
\begin{eqnarray}\mathit
T_{los}(l)=T_{sim}\left(\frac{\rho_{los}}{\rho_{sim}}\right)^{0.2}\\
n_{HI,los}(l)=n_{HI,sim}\left(\frac{T_{sim}}{T_{los}}\right)^{0.7}\left(\frac{\rho_{los}}{\rho_{sim}}\right),
\end{eqnarray}
where $\rho$ is the mass density, $T$ is the temperature, $n_{HI}$ is the neutral fraction of hydrogen and $\delta_{L}$ is a small-scale over-density. The subscript ``los'' refers to small-scale structure along the line of sight and the subscript ``sim'' again refers to values obtained directly from the simulation (low spatial resolution). 

The last two equations assume ionization equilibrium, which is a relatively good approximation after the overlap of HII regions (Gnedin 2000), and a density-temperature relation in the form $T\sim\rho^{0.2}$, which is a reasonable approximation for $5<z<6$ redshift interval (Hui \& Gnedin 1998, Gnedin 2001). 

The resulting distributions of density, neutral fraction, temperature and velocity, yield a synthetic absorption flux along the line of sight that includes the small scale fluctuations resulting in a synthetic spectrum that represents both large and small scale structure.  

\section{Quasars and HII Regions}
\label{sec-quasars}
Quasars are the brightest objects known at many wavelengths and thus they can be observed at high redshifts. The brightest quasars have a bolometric luminosity at z=6 of around $10^{12.5}L_{\odot}$. The emitted radiation is mainly in the UV, and can ionize a large region of gas surrounding the quasar, called the HII region.

In general the size of the HII region depends on both density and clumpiness of the gas and also the abundance of ionizing photons. This means that in addition to density and clumpiness, the age and luminosity of the source plays an important role in the process of forming the HII region and determine its spatial growth. 

Analyzing the distribution of the HII region sizes and how they evolve with time can give us information about the characteristics of quasars and the properties of the gas surrounding them. In this paper we use the lines of sight described in Section~\ref{sec-los} to determine the size of the HII regions. The edge of the \HII\ region around a quasar should be detectable as  a clear increase in Ly-$\alpha$ absorption just redward of the quasars Ly-$\alpha$ emission line. 
 
The fact that the neutral fraction increases further away from the quasar due to the dilution of photons can be used to determine the size of the HII region. It means that there will be more neutral gas further away from the quasar, so the ionizing radiation cannot escape and there is a decrease in transmitted flux. We define the edge of the HII region to be the minimum of the flux, before it increases again due to general reionization of the universe. To be able to measure this HII region in the simulation, the quasar needs to be bright enough to ionize an area large enough around itself so that the resolution of the simulation is not larger than the HII region. 

Additional problems in detecting the HII regions and determining their sizes are caused by the fact that the HII regions are not necessarily spherical and also that the quasars do not have to be positioned at their center. The irregular shape of the HII regions is mainly due to the ``lumpiness'' of the gas, which means that less dense parts of the IGM\footnote{The asymmetric shapes can clearly be seen in Figure~\ref{fig:quasarall} for illustration of this phenomenon.} are ionized faster than other more dense ones.

There are of course many different parameters that can influence the size distribution, only some of which can be understood and modeled. For example, larger regions can be caused by lower density in a relatively uniform IGM, higher luminosity of the quasars or lifetimes long enough for the quasars to reach equilibrium. 

The synthetic spectra can serve as the basis of our analysis of the effects of the different clumping factor models and the simulations of reionization. The following sections will show how we used these lines of sight to obtain distributions of \HII\ region sizes and other characteristics of the spectra.

\section{Results}
\label{sec-results}

\subsection{The mean transmitted flux}
\label{subsec-simulations}
Our primary goal is to reproduce the spectra of high redshift SDSS quasars as closely as possible. To accomplish this, the first step was to create a simulation of reionization covering a large volume and compare it to observations taken at high redshifts using SDSS data to better reproduce the mean transmitted flux. We ran a large number of simulations with varying $f_{esc}$ and clumping factor models to obtain the best fit to the SDSS data possible. These trial runs constrained $f_{esc}$ for each clumping factor model. Then this set of runs was used to test our approach and to determine whether our analysis with synthesized spectra was appropriate. Thus, we obtained several $64^{3}$ cell test runs with either $10\hh\dim{Mpc}$ or $2\hh\dim{Mpc}$ cell-size. 

The production runs have $128^{3}$ cells with again either
$10\hh\dim{Mpc}$ or $2\hh\dim{Mpc}$ cell-size for all three clumping
factor models. In addition we have runs for varying quasar lifetime
also with $128^{3}$ cells and $2\hh\dim{Mpc}$ cell-size, since this
combination of simulation parameters gives us good resolution and a
large enough volume to contain bright enough quasars. 

\hide{Quasars bright enough to be included in the SDSS are rare: there should be only about 100 such quasars at $z>6$ in our Universe. On average less than one SDSS quasar should be produced in each $(64 \hh\dim{Mpc})^{3}$ simulation (Schirber and Bullock, 2001). This means that the brightest quasars found in our simulation are not as bright as the ones observed by SDSS unless we increase the volume or increase the quasar number density at the high-luminosity end of the luminosity function. Therefore an increase in box size which allows us to cover a much larger volume, can give us a better chance of forming a SDSS size quasar in our simulation.}

\begin{figure}[!h]
\begin{center}
\epsscale{1.0}
\plotone{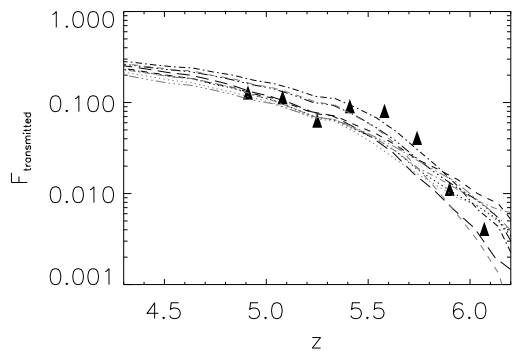}
\end{center}
\caption{Average flux versus redshift for the SDSS data and some of the best fits: SDSS data from White et al.(\textit{black individual triangles}), Run 3 (\textit{black long dashed}), Run 4 (\textit{gray dotted}), Run 5 (\textit{black short dashed}), Run 6 (\textit{gray dash dot dot dot}), Run 7 (\textit{gray dashed}), Run 8 (\textit{gray dash-dot}), Run 9 (\textit{black dots}), Run 10 (\textit{black dash dot}), Run 11 (\textit{black dash dot dot dot}). Notice how similar all models and resolutions appear when fitted with $f_{esc}$.}
\label{fig:fluxcurvesall}
\end{figure}

Table~\ref{tbl-runsum} summarizes a few of the best fit runs with all their initial parameters. Figure~\ref{fig:fluxcurvesall} compares, for each of these runs, the mean transmitted flux averaged in redshift bins $\Delta z=0.1$ from $z=4$ to $6$, to that measured by Songaila et. al (2004) from SDSS quasars. In all cases shown in Figure~\ref{fig:fluxcurvesall}, the mean transmitted flux gives a reasonable fit to the SDSS data. This means that for each model, the escape fraction can be adjusted to achieve consistency with the mean transmitted flux measured from the SDSS data. Further analysis, discussed in Section~\ref{subsec-HIIsize} and ~\ref{subsec-troughs}), reveals some difference between the cases, such as in the distribution of \HII\ sizes and of troughs in the spectra.

\begin{deluxetable}{ccccccc} 
\tabletypesize{\scriptsize}
\tablecaption{Summary of runs \label{tbl-runsum}}
\tablewidth{0pt}
\tablehead{\colhead{Run \#} & \colhead{case} & \colhead{box size} & \colhead{number of cells} & \colhead{cell size in $\hh\dim{Mpc}$} & \colhead{$f_{esc}$} & \colhead{quasar life-time}}
\startdata
1 &A &256 &128 &2 &5.0  &$\infty$ \\ 
2 &A &1280 &128 &10 &1.5  &$\infty$ \\
3 &B &128 &64 &2  &0.07  &$\infty$ \\ 
4 &B &640 &64 &10  &0.045  &$\infty$ \\ 
5 &B &256 &128 &2  &0.07 &$\infty$ \\ 
6 &B &1280 &128 &10 &0.035  &$\infty$ \\ 
7 &B &256 &128 &2 &0.014  &$t=10^{7}$ \\
8 &B &256 &128 &2 &0.079  &$t=10^{8}$\\
9 &B &256 &128 &2 &0.065 &$t=10^{9}$ \\
10 &C &256 &128 &2 &0.45 &$\infty$  \\
11 &C &1280 &128 &10 &0.35  &$\infty$ \\
\enddata
\end{deluxetable}

When comparing the different cases, we observe, as expected, that Case A has the largest $f_{esc}$. This is because we take into account all the photons in the high density regions. The best fit values for Case B are the smallest, only around a few percent (Section~\ref{sec-results}). The significant difference in the values for $f_{esc}$ for the various clumping factor models show how different the methods of counting photons are and how they affect the amount of ionizing flux, but the fact that we can fit all three models within reasonable error to the SDSS data shows that $f_{esc}$ is a convenient phenomenological parameter and not the same as the observationally measured quantity.

Fitting the effective escape fraction to get the correct shape in a flux versus redshift plot requires several simulation runs with varying escape fraction (see Section~\ref{subsec-simulations}) to close in on the correct $f_{esc}$. We mainly fit the average transmitted flux between $z=5$ and $z=6$, because of the larger errors on the SDSS data at higher redshifts and our focus on simulating the period of reionization.

In addition to the three different cases for the clumping factors, the best-fit values for the effective escape fraction also depend weakly on numerical resolution and box size. These effects are usually not large, but need to be taken into account to obtain the best fits to the data.

Figure~\ref{fig:fluxcurvesall} shows that the case of $10^{9}\dim{yr}$ quasar lifetime is barely distinguishable from that of infinite lifetime. In effect, $10^{9}\dim{yrs}$ is tantamount to infinite lifetime. The best fit escape fractions are similar: $f_{esc}=0.065$ for $10^{9}\dim{yrs}$, versus $f_{esc}=0.07$ for infinite lifetime. The effective escape fraction for the shorter lifetime is slightly smaller, because, for a given luminosity function, quasars with shorter lifetimes are less biased.

The normalized transmitted flux of all the runs shown in Figure~\ref{fig:fluxcurvesall} tends to $0.2$ at $z=4.5$ at the low redshift end, but for the high-redshift end there is a larger variation in the flux between the different runs. The data obtained from SDSS at redshifts $z>5$ have a larger error because there are fewer high redshift quasar observations, so that our results lie within the error for this regime. The general trend of the SDSS data is fit well, though the observed ``bump'' in intensity around $z=5.5$ does not appear clearly in any of our models. 

Figure~\ref{fig:fluxcurvesall} illustrates that the model for the clumping factors only weakly effects the evolution of the IGM during reionization. One of the slight deviations can be seen for Case C, where the high density regions are excluded from the calculation of the clumping factors. The mean transmitted flux curve shows a somewhat more rapid ionization around $z=5.5$ to $z=6.0$. The case C models follow the curve set by the SDSS data more closely, but at lower redshifts they tend to lower neutral fraction than case B models. Figure~\ref{fig:fluxcurvesall} emphasizes that independent of the choice of cell-size, box size and model, there is a $f_{esc}$ that fits the transmitted flux of the simulation run to the SDSS data. 

\begin{figure}[!h]
\begin{center}
\epsscale{1.0}
\plotone{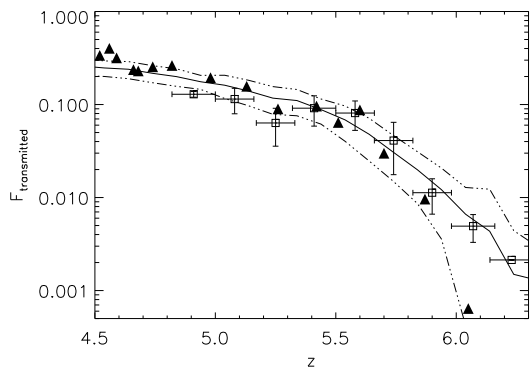}
\end{center}
\caption{Mean transmitted flux for two different analyses of the same set of SDSS quasars: Songaila et al. (2004) (\textit{black triangles}) and White et al. (2003) (\textit{open squares}) compared to the mean transmitted flux calculated for run 10 (which we choose just for illustration purposes; other models have very similar variations in the mean transmitted flux with redshift), a Case C $128^3$ simulation with $2\hh\dim{Mpc}$ cell-size (\textit{black line}). Shown also are the lines corresponding to $1 \sigma$ change at each redshift for the synthetic spectra (\textit{dash-dot-dot-dot}).}
\label{fig:sigmaplot}
\end{figure}

Figure~\ref{fig:sigmaplot} shows two sets of the same data analyzed independently by White et al. 2003 and Songaila et al. 2004. For clarity only the error bars on the White et al. set are displayed. Also shown is a mean transmitted flux curve from one of our simulation and the $1 \sigma$ deviation of the flux from this mean in the various lines of sight. The fit clearly lies within the error of the observational data. The variations in mean transmitted flux from the simulation increase toward higher redshift, indicating the spread in the timing of reionization along different lines of sight.

\begin{figure}[h]
\begin{center}
\epsscale{1.0}
\plotone{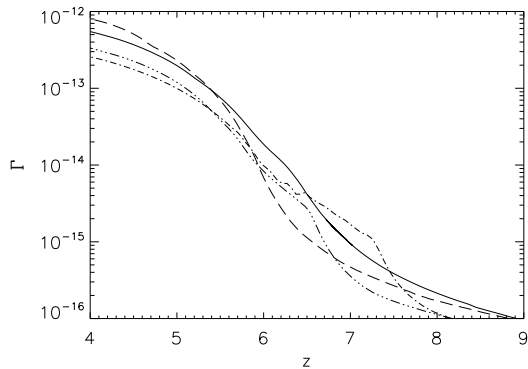}
\end{center}
\caption{Comparison of the averaged photoionization rate versus redshift for case B: Run 5(\textit{solid}), Run 6 (\textit{dash-dot}), Run 4 (\textit{dash-dot-dot}), and Run 3 (\textit{dashed}).}
\label{fig:tdsrecB}
\end{figure}

For a clearer illustration of how reionization proceeds in our simulation, Figures~\ref{fig:tdsrecB} shows the evolution with redshift of the averaged photoionization rate $\Gamma$ in different simulations. For these figures, the photoionization rate $\Gamma$ of the gas was averaged over the entire simulation box instead of along a line of sight. In all the different cases shown the rapid increase of $\Gamma$ at $z\sim 7$ is clearly visible, also the rate of increase of $\Gamma$ versus redshift is lower than predicted by small box simulations (Gnedin 2000, 2004). This corresponds to a drop in neutral gas which is spread over a range in redshift of only $\Delta z \sim 1$. 

Figure~\ref{fig:tdsrecB} illustrates how box and cell size affect the averaged photoionization rate $\Gamma$ for case B. The graph indicates a spread of about a factor of three while the ionization rate increases by $\sim 10^4$ as the redshift varies from $z \sim 9$ to $4$. The main systematic difference appears to be that the photoionization rate is on the whole (except during reionization) lower in the lower resolution simulations, with $10 \hh \dim{Mpc}$ cell-size. This difference can be attributed to the difference in clumping factors, and can be interpreted as an indication of the uncertainty arising from the use of the clumping factor formalism.

\subsection{Examples of Synthetic Spectra}
\label{subsec-spectraobs}

As described in Section~\ref{sec-los}, we use synthetic spectra to allow comparison between the results of our simulations and observations. Small scale variations are added to the synthetic spectra created from the simulations as described in Section~\ref{sec-los}.  Thus regions with low flux correspond to regions where the mean free path is too small for any significant amount of radiation to be transmitted. We also observe spikes in the spectra, that originate in the lower density regions. In the following paragraphs we compare the large-scale variations with the synthetic spectrafrom the convolution of large and small scales.

\begin{figure}[!h]
\begin{center}
\epsscale{1}
\plotone{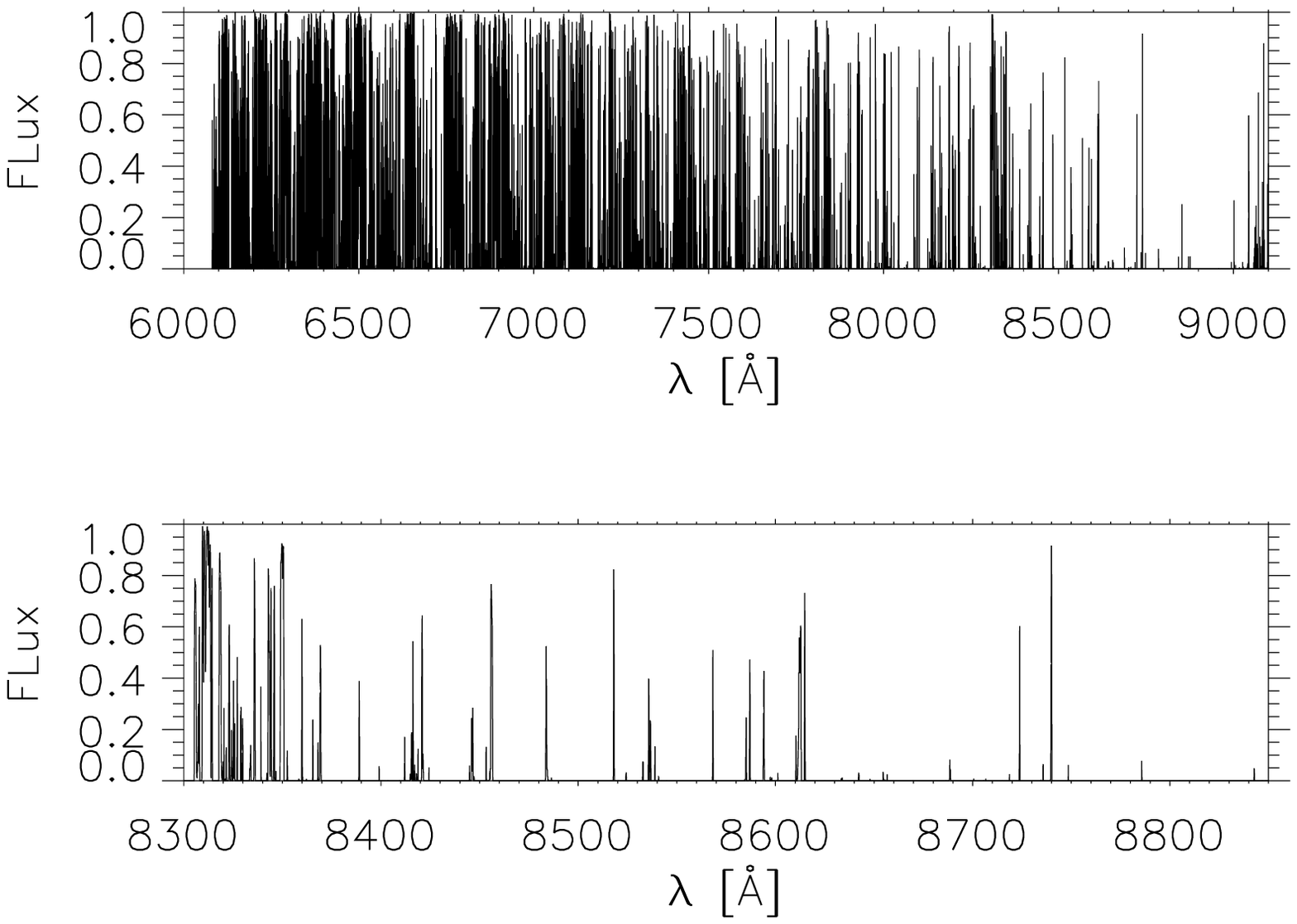}
\plotone{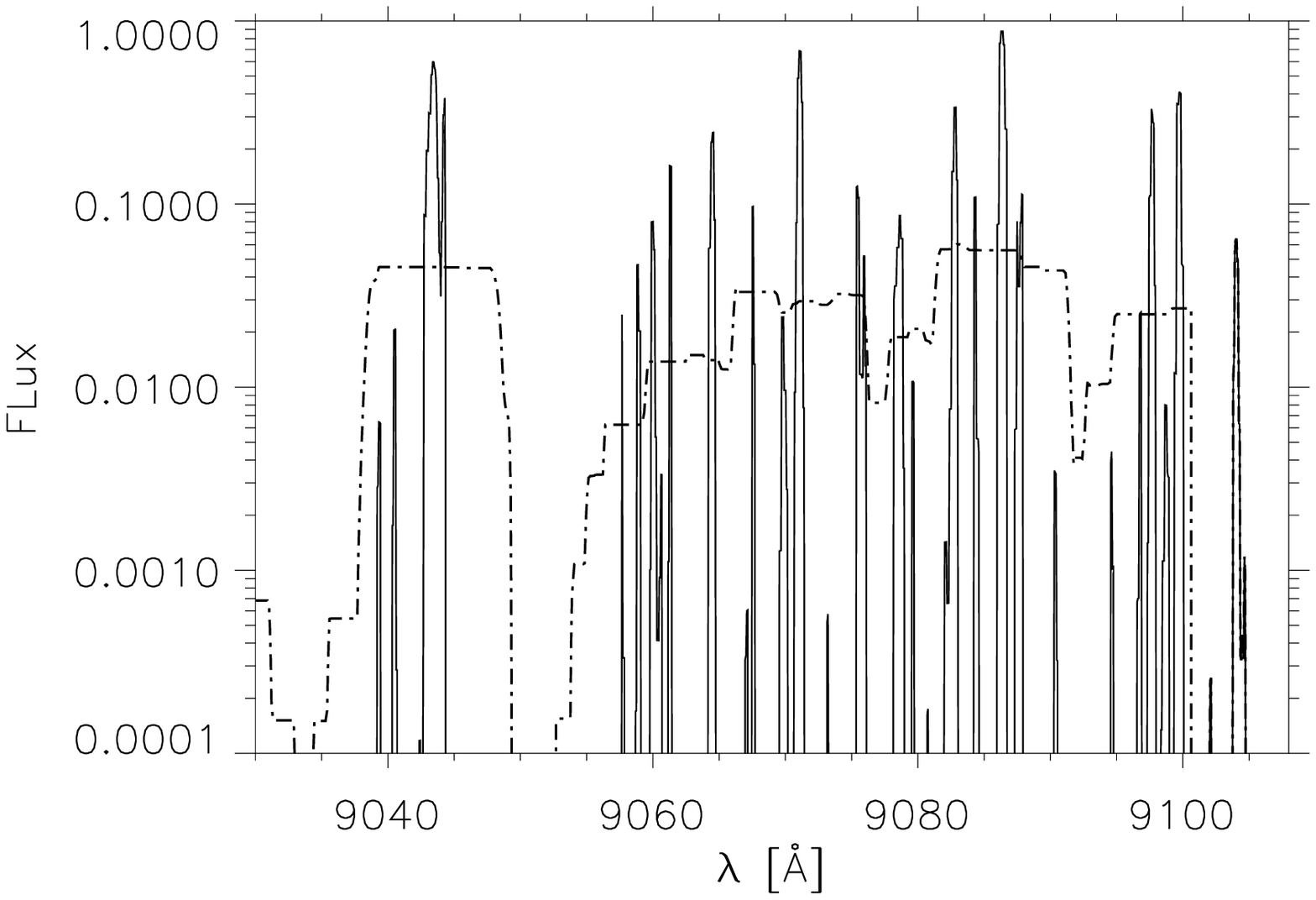}
\end{center}
\caption{The \textit{top panel} shows a synthetic spectrum of a high-redshift quasar at $z=6.49$ with the flux distribution from the small scale structure shown in \textit{black} and the flux from the large scale structure shown in \textit{gray}. The \textit{middle panel} shows a zoom to the troughs in the same spectrum. Note the dense absorption forest at these wavelengths and large troughs with no transmitted flux. The \textit{bottom panel} shows the same spectrum  as above but zoomed to the region near the quasar at $z=6.49$, the \textit{black} line shows the flux with small scale structure, the \textit{dash dotted} line shows the flux  smoothed over $10 \dim{\AA}$ to illustrate the measurement of an \HII\ region size.}
\label{fig:HIIspectrumlargesmall}
\end{figure}

Figure~\ref{fig:HIIspectrumlargesmall} displays both the spectrum with the small scale flux added in and the large scale fluctuations directly from the simulation over the complete wavelength range (\textit{top}) and near a bright quasar at $z=6.5$ (\textit{bottom}) and thus illustrates some of the properties of the synthetic spectra and their implications. The \HII\ region around the quasar at the bottom of Figure~\ref{fig:HIIspectrumlargesmall} is indicated by the large scale flux distribution (\textit{gray}). The limited resolution of the simulation, $\lambda=8.5\dim{\AA}$, smooths the flux distribution so that the \HII\ region is clearly delineated. Note that using the small scale distribution of flux to determine the HII region size would be more difficult, because the edge of the \HII\ region is not clearly defined. The resolution of the spectrum with the small scale structure is taken to be that of the Keck high-resolution instrument. The Lyman-$\alpha$-transmission is relatively sparse, which shows that the IGM is not ionized enough at this epoch to be transparent. This does not imply that the IGM is neutral, because even a small fraction of neutral hydrogen increases the optical depth drastically.

The top panel of Figure~\ref{fig:HIIspectrumlargesmall} shows another synthetic spectrum of a quasar at $z=6.49$, which is about the same distance as the furthest quasar observed so far, over the whole wavelength range from $9100\dim{\AA}$ to $6000\dim{\AA}$. \hide{For comparison, the resolution of this synthetic spectrum is the same as for spectra obtained with the best ground-based telescopes such as Keck.} This plot shows how the large-scale neutral fraction decreases with shorter wavelengths because the universe is more ionized (Miralda-Escud\'{e} et al., 2000) so that the average flux increases around $7500\dim{\AA}$ or $z\sim 5.2$, as can be seen in the large scale flux distribution and has been observed in SDSS spectra. Also, this figure shows how the lines thin out toward higher redshifts and the overall average flux decreases, similarly to the Gunn-Peterson trough found in observational spectra. 

The middle panel of Figure~\ref{fig:HIIspectrumlargesmall} shows the same spectrum as in the top panel. Here we zoom into some of the large gap regions centered at $8550\dim{\AA}$ and its surroundings to show the sparse transmitted flux in more detail. Gaps this large at a relatively low redshift should correspond to large-scale overdensitis in the IGM (see also Section~\ref{subsec-troughs}). It is also notable how dense the Lyman-$\alpha$ forest is at these redshifts. 

The bottom panel of Figure~\ref{fig:HIIspectrumlargesmall} is a zoom to the high redshift end of the spectrum. It shows the transmitted flux from the \HII\ region of the quasar both on small scales (\textit{black line}) and smoothed to a resolution of $10\dim{\AA}$ for calculating the \HII\ region size (\textit{dash-dot}).
 
\subsection{Analysis of Simulation for HII Regions Sizes}
\label{subsec-HIIsize}

A bright quasar can photoionize a large region around it, its \HII\ region. The \HII\ region is detectable in spectra as an increase in transmitted Lyman-$\alpha$ flux to the red side of the quasar's Lyman-$\alpha$ emission line.
Our spectra can be used to obtain properties of \HII\ regions and then compare them to observed properties. To analyze the sizes of the \HII\ regions, we calculate several characteristics of the HII regions surrounding a quasar and how they change with time. Some of these characteristics are also measurable in observed high resolution spectra: e.g. the maximum flux, the distance at which the flux is minimal and the area under the flux versus wavelength curve. These values give a measure of the size of the \HII\ region, and can be compared to observed data. Since flux increases on average with decreasing redshift due to the ionization of the IGM, the flux minimum should be a good measure of the boundary of the HII region.     

\begin{figure}[!h]
\begin{center}
\epsscale{1}
\plotone{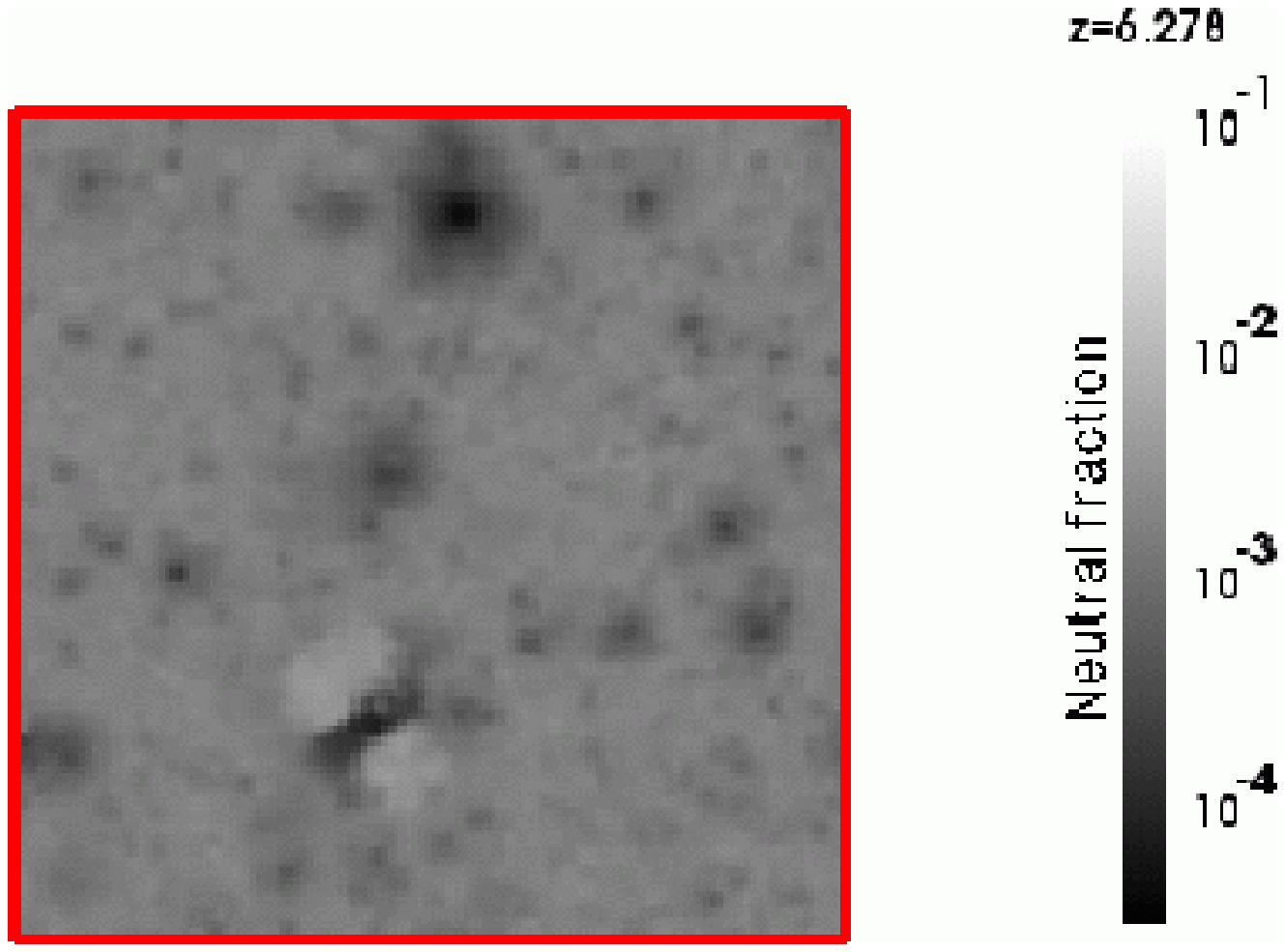}
\plotone{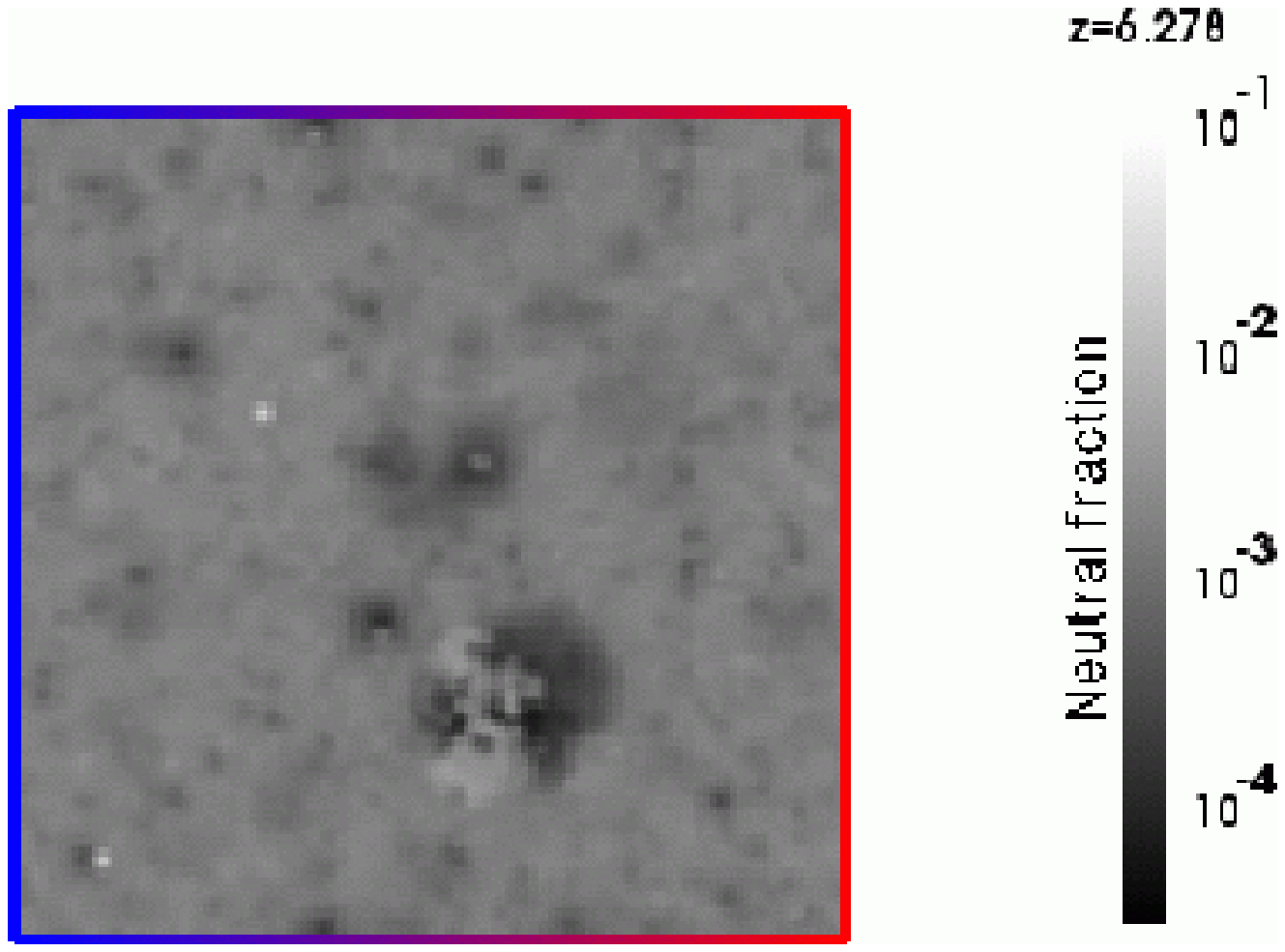}
\end{center}
\caption{{\it Top panel:\/} This x-y slice through the simulation box shows a view of the neutral fraction distribution in a $128^{3}$ simulation with $2h^{-1}Mpc$ cell-size at $z=6.28$. Several luminous quasars and their extended \HII\ regions can be see as darker spots. For example the irregularly shaped dark region in the lower left corresponds to a \HII\ region of a luminous quasar surrounded by dense more neutral gas (shown in \textit{light gray}). {\it Bottom panel:\/}
A different view (z-y plane) of the simulation, with the same bright quasar in the bottom right corner. Note the asymmetry in the shape of the \HII\ region (\textit{dark gray}).}
\label{fig:quasarall}
\end{figure}

Small scale fluctuations and low spatial resolution can prevent us from finding the actual minimum. Also our simulation does not produce spherical HII regions, as can be seen in in the upper two panels of Figure~\ref{fig:quasarall}, causing several different values for the size of an HII region depending on the direction of the line of sight cast. The images in Figure~\ref{fig:quasarall} show two cross-sections of the simulation box gray-scale coded by the neutral fraction from $x_{HI}=10^{-5}$ in black to $x_{HI}=10^{-1}$ in white ($x_{HI}=\frac{n_{HI}}{n_{H}}$). The top cross-section lies in the x-y plane, while the bottom one shows a z-y projection. The slices are positioned at the position of a luminous quasar in the simulation and shows the large \HII\ regions ($\textit{black}$) surrounding it as cross-section through the simulation box. 

In the top image of Figure~\ref{fig:quasarall}, one can clearly see a relatively spherical \HII\ region surrounding a bright quasar. Near this large \HII\ region we can observe other smaller highly ionized regions in dark gray, corresponding to $x_{HI}<10^{-4}$. These regions correspond to fainter quasars and their HII regions. In the lower right part of the image is an example of a clearly non-spherical \HII\ region around another very bright quasar. It is more difficult to obtain a reliable value for the size of the \HII\ region because it is so irregularly shaped. Since most of our quasars produce ionized regions that are non-spherical, we cast several lines of sight from each quasar to obtain a statistically relevant sample.

The bottom panel of Figure~\ref{fig:quasarall} shows a perpendicular slice of the same simulation. Here the complicated structure of the \HII\ region is even more pronounced with features that resemble a plume.  There are also a few white spots in the image, which correspond to highly neutral regions in the simulation. These spots would show up as regions in spectra with no transmitted flux. 

\begin{figure}[h!]
\begin{center}
\plotone{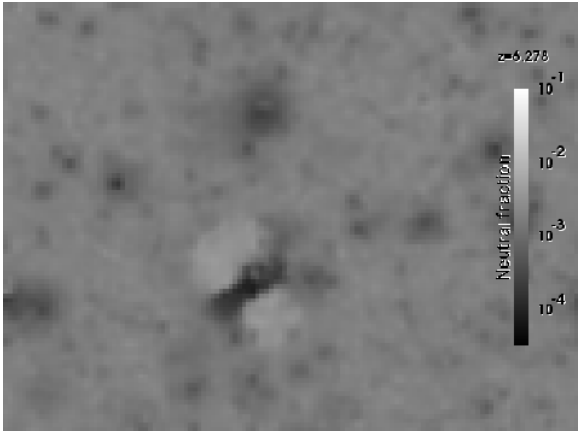}
\end{center}
\caption{A close-up view of the \HII\ region of the bright quasar in the simulation from Figure~\ref{fig:quasarall}. Again, the asymmetry in the shape of the \HII\ region is clearly visible(\textit{dark gray})}
\label{fig:quasarclose}
\end{figure}

Figure~\ref{fig:quasarclose} gives a closer view of the same HII region. The black area is the highly ionized region and has a diameter of about $8\hh\dim{Mpc}$, however the ionized region extends much further (shown in darker gray). As mentioned above, we can see the irregular, elongated shape of the ionized region, surrounded by almost completely neutral gas on either side. This high density gas has not yet been fully ionized and confines the \HII\ region to a ``tunnel''. The ionizing radiation can escape from the quasar only through this tunnel, because the mean free path is much lower in the orthogonal direction. We can again see several areas with low neutral fraction around the main HII region. Most of these spots that have a neutral fraction of about $10^{-3.5}$ are clustered together with other more ionized ``blobs'' shown in lighter gray.

At this redshift ($z=6.28$) the universe is already very close to fully ionized, so the dark gray areas that have a neutral fraction of $10^{-3}-10^{-4}$ are the most neutral regions in the simulation at this redshift. However this does not imply that the spectra will have a large transmitted flux, because even a neutral fraction as low as $10^{-4}$ is sufficient to absorb all flux from the quasar. 

\hide{
In addition to the irregular shapes of the ionized regions, having most of the universe ionized to such a high degree complicates finding accurate sizes for the HII regions, because the difference in transmitted flux through a region can be too small to be measured. This effect appears as reduced intensity at the high redshift end of the spectrum.}

\subsection{Properties of HII Regions}

 The sizes of \HII\ regions measured from the extent in wavelength of the region of increased flux offers a point of comparison between simulations and observations. 

The simulations show that a quasar's \HII\ region is typically neither spherical nor sharply defined, so the notion of its size is inevitably somewhat hazy. We define the size of an \HII\ region for our synthetic spectra as the first minimum with $F<0.01$ or $F<0.03$. The flux cutoff is calculated by smoothing the synthetic spectra including the small scale flux, so that the resolution is $10\dim{\AA}$. The first minimum of the flux that falls below the cutoff flux is defined as the edge of the \HII\ region and thus corresponds to a region where there is no or little flux on a scale of $10 \dim{\AA}$. 

This definition reflects the fact that the degree of ionization should be high in the immediate vicinity of the quasar, should decrease moving away from the quasar, and then should increase again because of the general ionization of the universe. It is also similar to methods for determining the size of \HII\ regions from observational spectra. 

The location of the minimum of the transmitted flux, which yields the size of the \HII\ region, depends on the resolution of the spectrum. To allow unbiased comparison between simulations and observations, the resolution of the synthetic spectra must match closely that of the observations. 

Because the \HII\ regions are aspherical, different lines of sight yield different measures of its size. For each simulation, we construct synthetic spectra from three random lines of sight through each of the fifty brightest quasars. The result is an ensemble of $\sim 150$ spectra from each simulation. It is this ensemble of spectra for each simulation which we compare to observations in section~\ref{sec-results}.  

\begin{figure}[!h]
\begin{center}
\epsscale{1}
\plotone{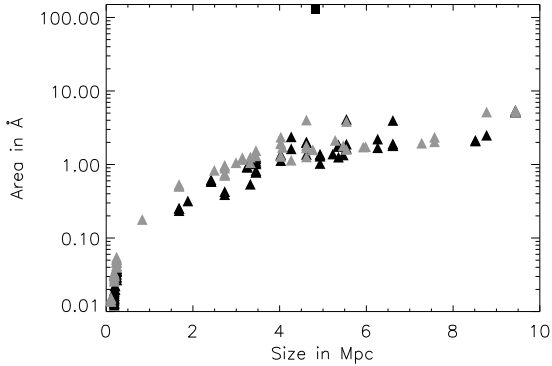}
\end{center}
\caption{Distribution of \HII\ regions in a $128^{3}$ simulation of case B, $2\hh\dim{Mpc}$ cell size and with infinite quasar lifetime (\textit{gray triangles} correspond to the cutoff at $F<0.03$ and \textit{black triangles} corresponds to $F<0.01$), also indicated is the \HII\ region size for the quasar SDSSJ10230+0524 (\textit{black square})}
\label{fig:HIImar31FAN}
\end{figure}

For example, Figure~\ref{fig:HIImar31FAN} shows one of the distributions of the size of the HII region as a function of the area under the flux versus wavelength curve. The area under the spectrum indicates the degree to which the gas in the \HII\ region is ionized. The distribution shown is for a simulation with case B clumping factors, $2 \hh\dim{Mpc}$ cell size and $128^3$ cells at $z=6.28$. In general the distance to the minimum flux and the area under the spectrum are two properties that can be observed and then compared to our simulations if the resolution is known. For example the data for SDSSJ1030+0524 is also included in the figure (\textit{black square}).

\begin{figure}[!h]
\begin{center}
\epsscale{1}
\plotone{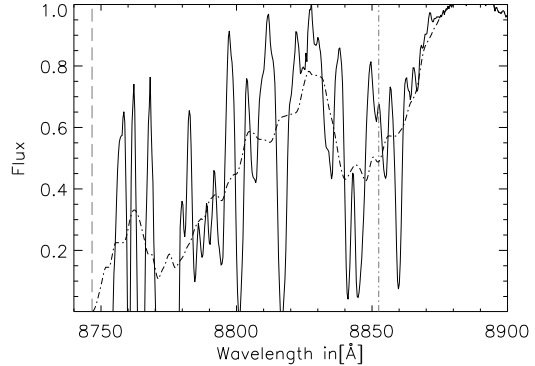}
\end{center}
\caption{Spectrum of SDSSJ1030+0524 (from X.Fan) in the immediate surrounding of the Lyman-$\alpha$ line. The gray curve illustrates the smoothed spectrum to a spatial resolution corresponding to a $2\hh\dim{Mpc}$ cell-size. Shown as vertical lines are the wavelength of Lyman-$\alpha$ (\textit{dash-dotted}) and two different measurements of the assumed edge of the \HII\ region: minimum with $F<0.01$ (\textit{dash-dot-dot-dot}) and minimum for higher resolution the smoothed flux (\textit{dashed}).}
\label{fig:HIIFAN}
\end{figure}

Figure~\ref{fig:HIIFAN} illustrates the measurement of the \HII\ region size of SDSS1030+0524 from the high-resolution spectrum. The gray curve displays the spectrum smoothed to a spatial resolution corresponding to $10\dim{\AA}$. The position of the redshifted Lyman-$\alpha$ line (\textit{dash-dot}) and two variations of determining the assumed edge of the \HII\ region (\textit{dashed} and \textit{dash-dot-dot-dot}) are shown as vertical lines. The line at $8748 \dim{\AA}$ shows the edge of the \HII\ region using the smoothed flux and finding the first minimum that also lays below the threshold flux of $0.01$. This measurement is the same as the method used on the simulation data and yields a size of $4.76 \dim{Mpc}$.

The discrepancy of the area under the spectrum, apparent from Figure~\ref{fig:HIImar31FAN}, for the quasars from our simulation and the SDSS quasar can be attributed to the much higher luminosity of the SDSS quasar. Higher luminosity quasars will emit more photons, which not only increases the size, but also reduces the ionization fraction of the gas in the \HII\ region. When comparing this graph to Figure~\ref{fig:HIIspectrumlargesmall} it is clear that the transmitted flux near the quasar, corresponding to the amount of ionized material, is significantly larger for the observed quasar than the transmitted flux for any of our simulated ones. This means that the gas is more highly ionized around this quasar.

\hide{ The fact that the \HII\ region is not larger in radius can be attributed to the amount of time that the quasar has been shining. The assumption is that the SDSS quasar has not yet had time to expand its \HII\ region further.}

\begin{figure}[!h]
\begin{center}
\epsscale{1}
\plotone{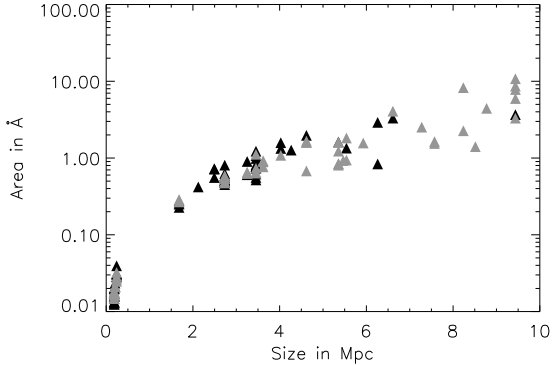}
\end{center}
\caption{Comparison of two distribution of \HII\ regions in $128^{3}$ simulations of case B with infinite quasar lifetime, one with $2\hh\dim{Mpc}$ cell-size(\textit{gray triangles}), the other one with $10\hh\dim{Mpc}$ cell-size(\textit{black triangles}).}
\label{fig:HIIresol}
\end{figure}

Figure~\ref{fig:HIIresol} displays the effects of simulation resolution on measuring the \HII\ region sizes by comparing $2\hh\dim{Mpc}$ and $10\hh\dim{Mpc}$ cell size runs. Two distributions are rather similar, also we find fewer \HII\ regions for lower resolution simulations, as would be expected.

\begin{figure}[!h]
\begin{center}
\epsscale{1}
\plotone{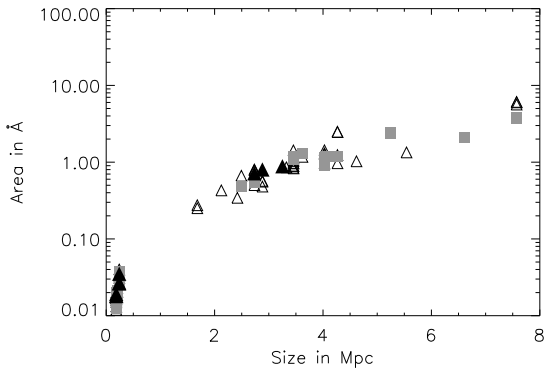}
\end{center}
\caption{Comparison of three distribution of \HII\ regions in $128^{3}$ simulations of case B with $2\hh\dim{Mpc}$ cell-sizeand with quasar lifetimes  of $10^9\dim{yrs}$ (\textit{open triangles}), $10^8\dim{yrs}$ (\textit{gray squares}) and $10^7\dim{yrs}$ (\textit{black triangles}).}
\label{fig:HIIalltcases}
\end{figure}

Figure~\ref{fig:HIIalltcases} shows the dependence of the size of the \HII\ regions on quasar lifetime. Of the three lifetimes shown, $10^7$, $10^8$ and $10^9$ years, the longest, $10^9$ years, gives results that are barely distinguishable from infinite lifetimes. For the shortest lifetime, however, one can see the effects of new quasars starting new \HII\ regions. For longer lifetimes the quasar can emit photons into the IGM for longer time periods after ionizing its immediate high-density surroundings, so that the \HII\ regions for shorter lifetimes will be smaller.

\hide{ Figure~\ref{fig:HIItime} displays the evolution with redshift of \HII\ region sizes for one simulation run. We obtained additional outputs, but they were omitted for clarity of the figure. Even though there is a significant scatter in the data, the HII regions grow in size and the area under the spectrum increases toward smaller redshifts. The difference between $z=6.5$ and $z=5.7$ is clearly visible. The distribution for the lower redshift only barely overlaps with the one from the higher redshift simulation. The area under the spectrum does not increase as much as the spatial extension of the \HII\ regions because the IGM is more ionized overall so that the ionizing photons can travel further before being absorbed.}

\hide{
\begin{figure}[!h]
\begin{center}
\epsscale{1}
\plotone{threezsbw.ps}
\end{center}
\caption{Distribution of \HII\ region sizes for three different redshifts: z=$6.5$ (\textit{filled gray squares}) , z=$5.9$ (\textit{open gray squares}), z=$5.7$ (\textit{filled black triangles}) all for infinite quasar lifetime}
\label{fig:HIItime}
\end{figure}}

\begin{figure}[!h]
\begin{center}
\epsscale{1}
\plotone{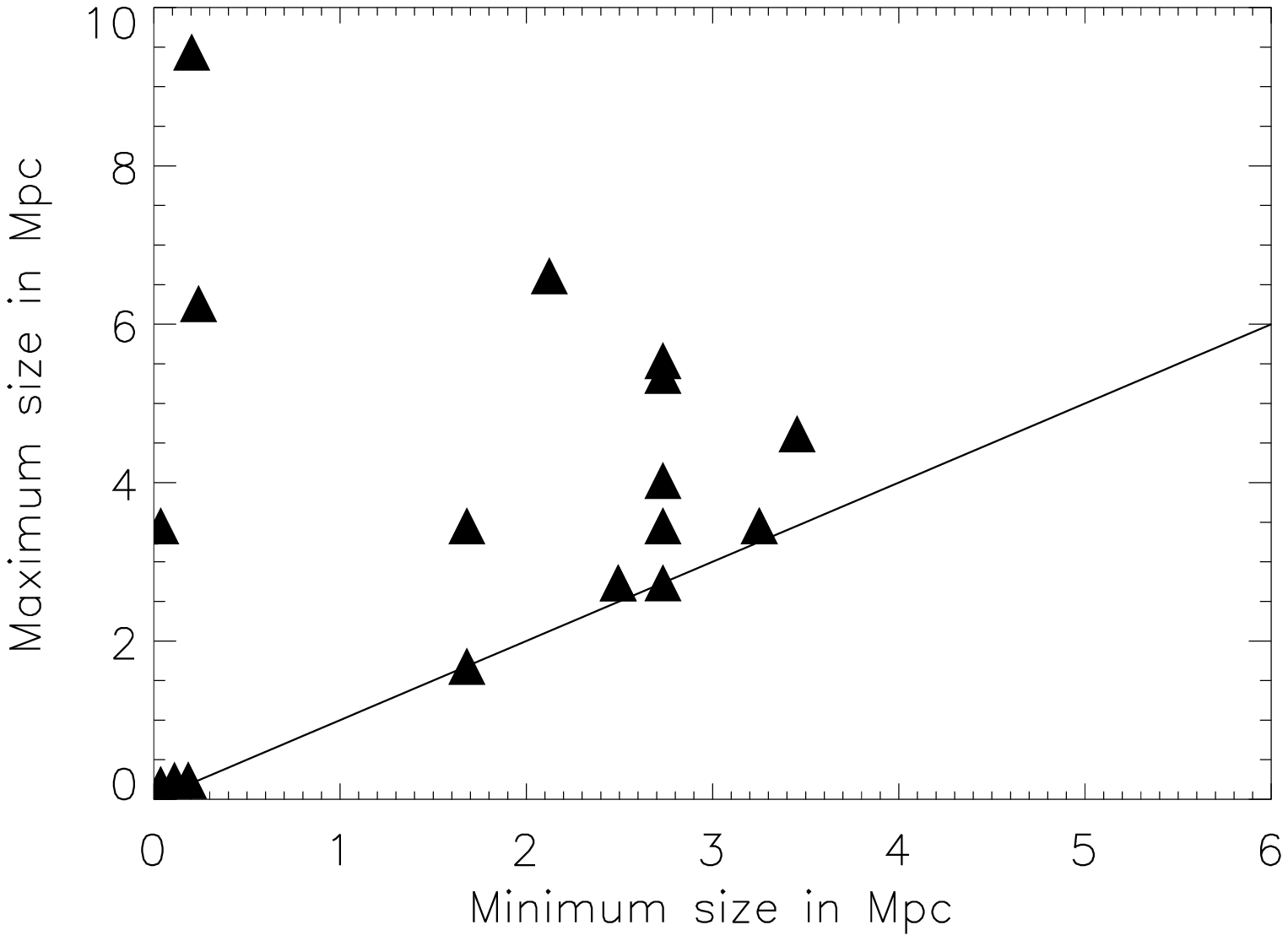}
\end{center}
\caption{This Figure shows the irregularity of \HII\ region sizes for a $128^3$ simulation with $2\hh\dim{Mpc}$ cell-size(\textit{black filled triangles}) with the maximum size versus the average size from three lines of sights. The line corresponds to spherical \HII\ regions. }
\label{fig:HIIirreg}
\end{figure}

The lines of sight for the spectra start at the position of the quasar and extend in a random direction through the box, therefore we only probe one specific part of each \HII\ region with each large-scale spectrum. Thus, using just one line of sight for measuring the size of the ionized region surrounding each quasar does not necessarily represent the actual \HII\ region size. Consequently, casting three lines of sight from each quasar should help lower the chance of not detecting an \HII\ region. We will cover different directions pointing out from the source so that even if the quasar is located at the edge of a very high density region in the IGM, we should be able to detect the HII region in one of the three lines of sight cast. We used each line to measure the size of the \HII\ region separately and so accumulated distributions for their sizes\footnote{Because most of the \HII\ regions are barely resolved even with $2\hh\dim{Mpc}$ resolution, using more than 3 lines of sight does not actually improve statistics.}.

Figure~\ref{fig:HIIirreg} addresses the question of shape of the \HII\ regions. The three lines of sight for each quasar yield a maximum and a minimum size, which are shown in \textit{filled triangles}. The line corresponds to spherical \HII\ regions, where the two parameters would be equal. The figure shows that almost none of the \HII\ regions are spherical. Some of the spectra show a large discrepancy between maximum and minimum size, for a few we do not even detect the \HII\ region in one of the directions within the limits of our numerical resolution. Thus, the measurement of \HII\ region sizes from one line of sight of a quasar does not provide a complete picture of the volume contained in the \HII\ region.

One of the difficulties in determining the \HII\ region size from this simulation is that our spatial resolution for the spectra can only be as small as the cell-size of the simulation box. This causes smoothing on scales of $2$ to $10\hh\dim{Mpc}$, depending on the simulation run. This means that we can easily miss or overestimate the HII region size. If the quasar is positioned in an extremely high density region, the ionized volume surrounding it can be too small to appear in a spectrum with such high resolution. Also, if the smoothing causes the edge of the HII region (set to be at the closest flux minimum to the quasar in concordance with the method used for the synthetic spectra) to disappear, the size of the ionized volume due to the quasar can be much smaller than determined by our method.

\subsection{Troughs in the Spectra}
\label{subsec-troughs}
In addition to analyzing the ionization of the IGM around luminous sources, we can also investigate the distribution of low flux regions, associated with higher neutral fraction, using the synthetic spectra. Regions in the simulation volume with low ionization fraction along the line of sight will appear as troughs in the absorption spectrum. The extent of these troughs in the spectra and their positions in space as determined by their redshifts, give us information about structure of the IGM between the quasar and the observer. Since only such a small amount of neutral gas is required to absorb almost all of the flux, the depth of the troughs is not used for analysis.

\begin{figure}[h]
\begin{center}
\epsscale{1}
\plotone{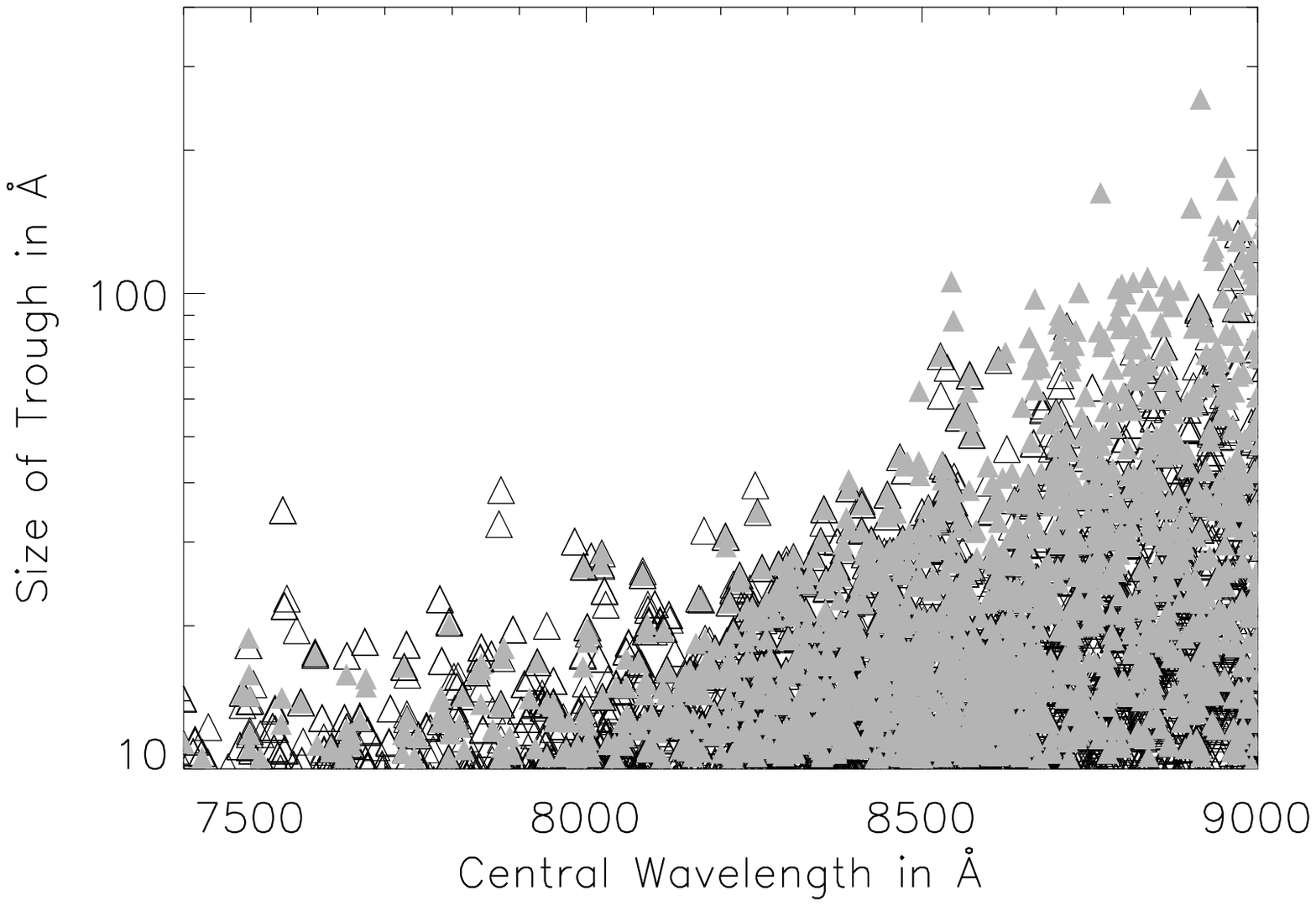}
\end{center}
\caption{Comparison of two distributions of troughs in 150 spectra for a $128^{3}$ simulation with $2h^{-1}Mpc$ cell-size at $z=6.5$, case B (\textit{black open triangles}) and case C (\textit{filled gray triangles})}
\label{fig:HIItroughscase}
\end{figure}

\begin{figure}[h]
\begin{center}
\epsscale{1}
\plotone{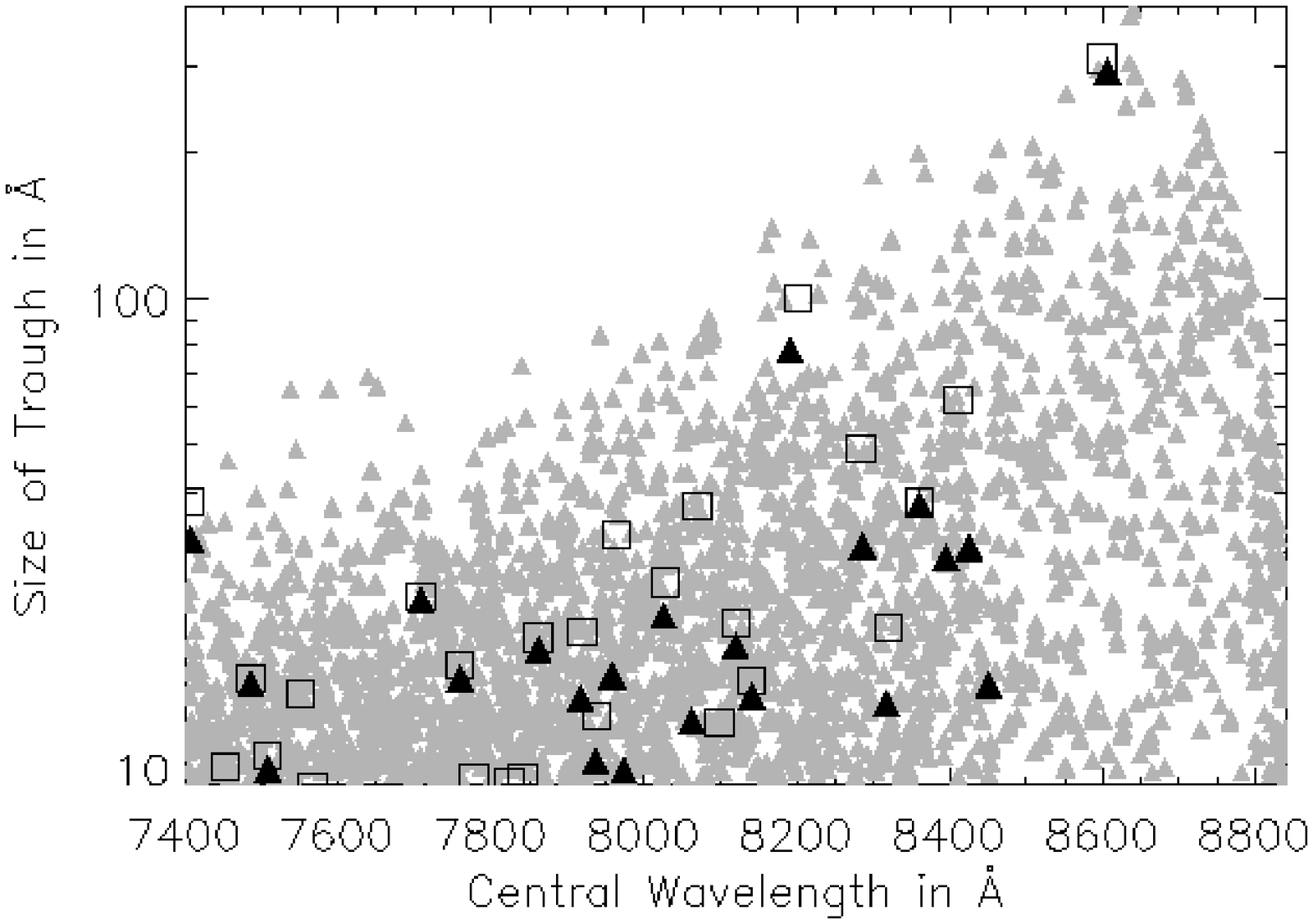}
\end{center}
\caption{Distribution of troughs for a $128^{3}$ simulation with $2h^{-1}Mpc$ cell-size at $z=6.28$ (\textit{small gray triangles}) with the troughs from SDSSJ1030+0524 overlayed  for a three $\sigma$ detection (\textit{black triangles}) and for a five $\sigma$ detection (\textit{open squares}). The simulation spectra were smoothed to the spectral resolution of observations and noise was added to better compare it to the observations.}
\label{fig:troughsmar31FAN}
\end{figure}

For this analysis, we use the same sets of synthesized spectra as described in Section~\ref{sec-los} and scan them for regions with extremely low flux (below a threshold detectable by instruments, here set to $10^{-3}$). We then compile distributions of these troughs from the 150 spectra obtained from each simulation and plot their distributions. A few examples of these distributions are shown in Figures~\ref{fig:HIItroughscase} through ~\ref{fig:troughsmar31FAN}. These figures show only the largest troughs, which extend more than $10\dim{\AA}$. 

We scanned the spectra for troughs between $\sim9100\dim{\AA}$ (corresponding to $z=6.5$) and $\sim7200\dim{\AA}$ (corresponding to $z=4.9$). The most striking feature of this set of distributions is their similarity. For example all the different runs display a decrease in average length of the low flux regions toward shorter wavelengths. This feature can be explained by the increase in the mean free path for ionizing photons due to the reionization of the universe. 

Figure~\ref{fig:HIItroughscase} illustrates the similarity of the distributions by taking a closer look at the trough distributions for two different clumping factor models (Case C in \textit{gray} and Case B in \textit{black}). The two distributions nearly completely overlay on each other. The only visible difference is a small excess of large high redshift troughs for case C and an excess for case B at lower redshifts ($\lambda\sim 7700$).

Another notable similarity is the fact that the largest troughs for each simulation are all about the same size, $100\dim{\AA}$. Most of these long stretches of low flux in the spectra lie above $8500\dim{\AA}$, with a few outliers in some of the simulations. For example Figure~\ref{fig:HIItroughscase} shows two troughs of about $40\dim{\AA}$ centered at $7800\dim{\AA}$ for case B, which corresponds to a stretch of $\sim 3.5\dim{Mpc}$ at redshift 5.4 or $14\hh\dim{Mpc}$ in comoving units. Considering that a trough only appears in our distribution when the flux is zero, these two troughs are produced by very large regions of relatively neutral gas in the IGM. For example, Figure~\ref{fig:HIIspectrumlargesmall} shows a spectrum containing a large trough near $z=6.0$ that extends for $100\dim{\AA}$, similar to the largest troughs shown in our distribution.

Figure~\ref{fig:troughsmar31FAN} shows that we can, similarly to the \HII\ regions size calculations, compare our results for trough distributions from synthetic spectra with the quasar spectrum SDSS J1030+0524. Since the data of the observed spectrum contains noise, we have to set a level of uncertainty below which we assume that the transmitted flux is zero. The \textit{black triangles} correspond to $3\sigma$ as the threshold for no transmitted flux and the \textit{open squares} correspond to a threshold of $5\sigma$. The distribution for the troughs obtained from the simulation is shown in Figure~\ref{fig:troughsmar31FAN} in \textit{gray triangles} and the similarities of the observation to the simulation are clearly visible. The $3\sigma$ detection limit yields a better agreement with our data, but both values show the same decrease in large troughs toward lowed redshifts and similar sizes for the troughs on the whole redshift range.

\begin{figure}[h]
\begin{center}
\epsscale{1}
\plotone{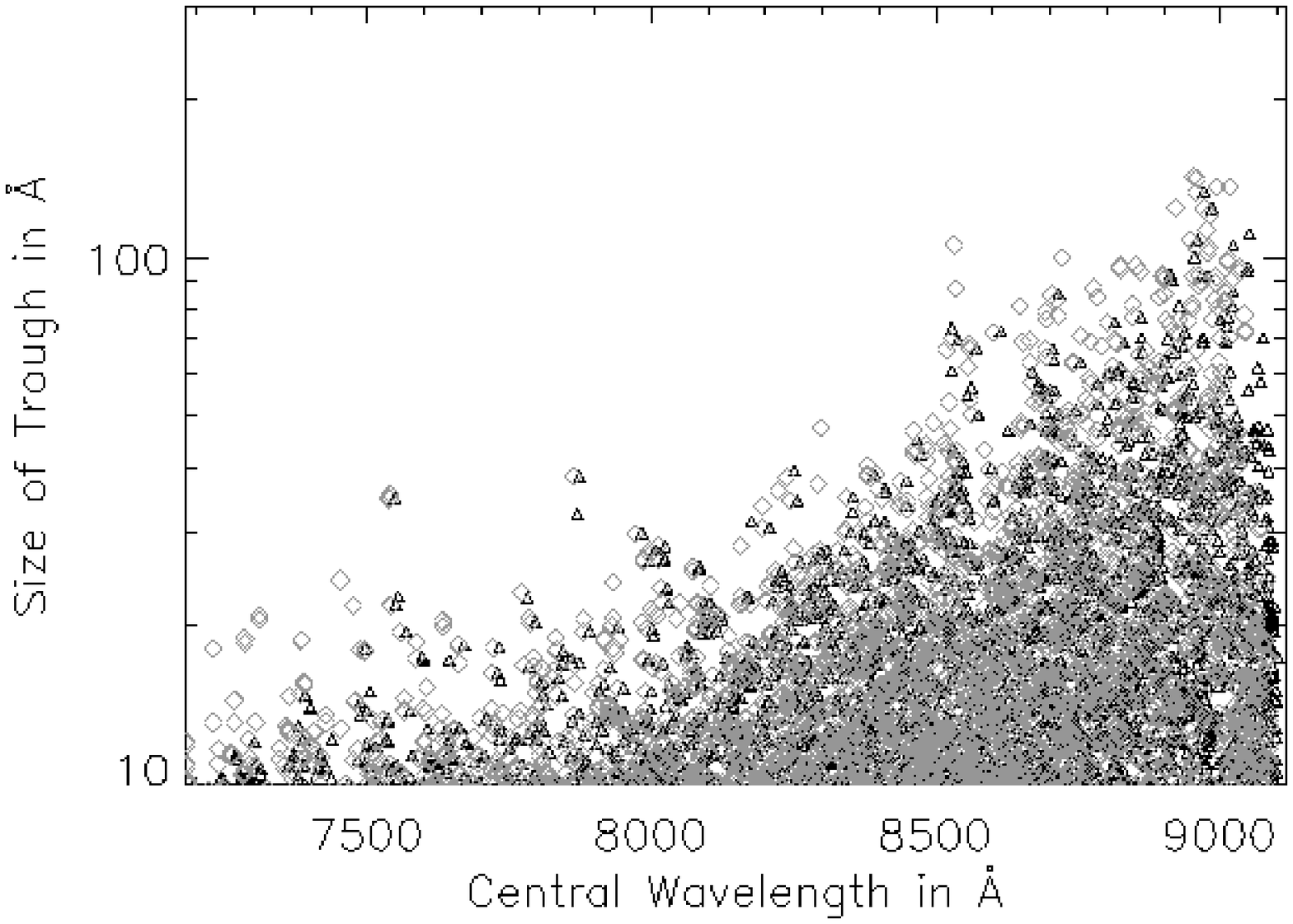}
\end{center}
\caption{Distribution of troughs for two $128^{3}$ simulations, one with $2\hh\dim{Mpc}$ cell-size (\textit{filled black triangles}) and one with $10\hh\dim{Mpc}$ (\textit{gray diamonds}). This comparison of resolutions shows no significant difference between the two cell sizes.}
\label{fig:troughsresolB}
\end{figure}

In contrast to measurements of HII region sizes, the comparison of resolution sizes of the simulation (Figure~\ref{fig:troughsresolB}) does not reveal a strong dependence on resolution. The largest trough size for both cases is around $100\dim{\AA}$ at high redshift and both show a slow decrease in trough size toward the lower redshifts. In addition they both exhibit a few outliers of larger size as mentioned above. There is a slight excess of larger troughs for the $10\hh\dim{Mpc}$ simulation. This is due to the larger step size in the synthetic spectra which removes some of the small spikes in transmitted flux that intersect large troughs. 

Altogether it is clear that the differences for varying parameters of the simulations do not change the distribution of troughs in the spectra significantly. This invariance to parameter changes allows us to conclude that as long as we fit the observational data from SDSS with the initial conditions we will receive the same information about the Lyman-$\alpha$ forest.

\subsection{Flux decrement}
\label{subsec-decrement}

\begin{figure}[h]
\begin{center}
\epsscale{1}
\plotone{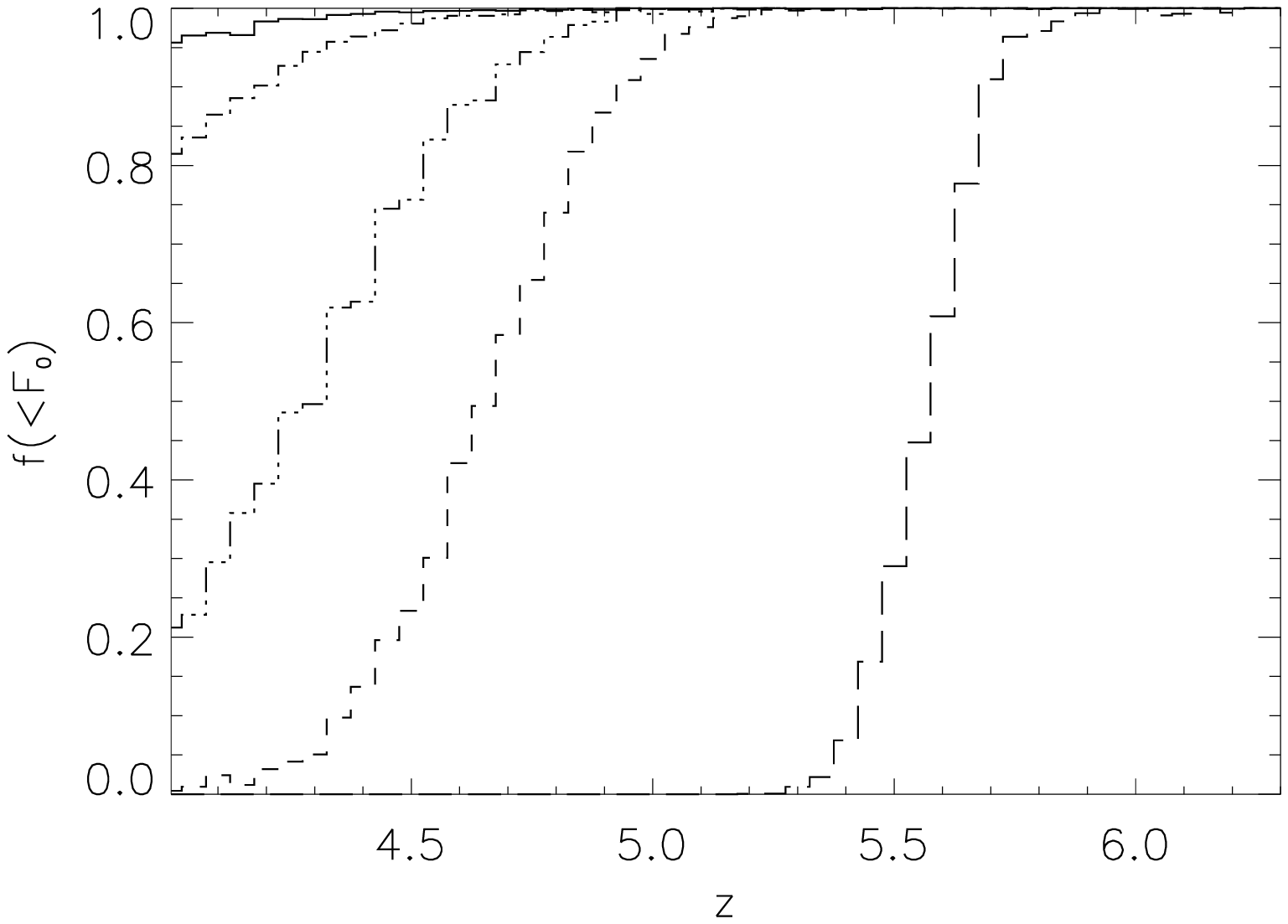}
\end{center}
\caption{Fraction of pixels with transmitted flux $F_{0}<0.3$ (\textit{continuous}), $F_{0}<0.2$ (\textit{dot-dashed}), $F_{0}<0.1$ (\textit{dot-dot-dashed}),  $F_{0}<0.05$ (\textit{short dashed}) and  $F_{0}<10^{-5}$ (\textit{long dashed}). The rapid increase in transmitted flux is apparent when the last percent of the neutral gas is ionized.}
\label{fig:fluxdecrement}
\end{figure}

In addition to looking at how the mean transmitted flux changes over time, the change in ionization of 
the IGM can also be observed when we consider how many pixels of the spectra fall below a certain flux
 at each redshift. Figure ~\ref{fig:fluxdecrement} shows this ``flux decrement'' for $5$ different
 levels: $F_{0}<0.3$, $F_{0}<0.2$, $F_{0}<0.1$, $F_{0}<0.05$ and $F_{0}<10^{-5}$. One can also observe 
that even at the low redshift end, around $z=4.0$, most of the pixels still have $F < 0.3$.
The curve corresponding to lowest flux limit, which reflects to the region in redshift where effectively all the flux is 
absorbed in neutral gas, displays a sharp boundary around $z=5.5$, below which all pixels have more 
flux than $10^{-5}$. This means that after redshift $5.5$, the universe is ionized enough to eliminate
the Gunn-Petersen trough. 

It is also important to note that all the distributions reach $1.0$ toward higher redshifts, meaning that there is 
not a significant amount of flux transmitted in this regime. Thus we can identify the region in the 
spectra above $z=5.5$ as the Gunn-Peterson trough. The change in transmitted flux appears to happen within 
a short period of time, supporting the idea that reionization happened quickly.

\section{Discussion and Conclusions}

Our cosmological simulations are used to model the large-scale effects of reionization including luminous high redshift quasars. A major issue with simulations of large volumes is that a large dynamic range needs to be covered including cosmic structures on a diverse range of scales. To solve this problem, we have used clumping factors to approximate the small scale structure of the cosmic gas. These clumping factors are derived from small volume simulations that can simulate structures on $\dim{kpc}$-scale and then used to produce results on $\dim{Gpc}$ scales with resolution up to $10h^{-1}Mpc$. We developed three models for computing the clumping factors: 1) case A, where clumping factors are weighted by volume, thus including local absorption and ionization, 2) case B, where the clumping factors are weighted by the inverse of the density therefore lowering the influence of high density regions and 3) case C, where we disregard all volume elements with high gas density and with that all absorption and ionization in high density regions. All three cases yield similar results and can fit the mean transmitted flux as observed by SDSS. 

 We use these simulations to construct synthetic spectra of quasars comparable to observed spectra of SDSS quasars at high redshifts. These synthetic spectra are used to investigate HII region sizes of bright quasars in our simulations and to compare with \HII\ regions of observed quasars. We found a significant scatter in the distribution for the sizes, even when looking at a large number of quasar HII regions. Our investigation shows that there is a dependence of HII region size on quasar lifetime (Figure~\ref{fig:HIIalltcases}), with larger HII regions for longer quasar lifetimes. However, this relation is too weak to put constraints of quasar lifetimes by observing their HII regions. 

The synthetic spectra can be used to make detailed comparison between the
observational data and the simulations. In this first, introductory paper, we
limit this comparison only to the distribution of \HII\ regions around bright
quasars and the distribution of throughs (regions with no flux) in the
spectra. In the future work we plan to use synthetic spectra for a more
detailed and elaborate comparison with the observational data.

Future improvements to our method will include, as has been mentioned in
\S~\ref{subsec-clump}, an account for the scatter in the modeled clumping
factors. As we have discussed above, the clumping factors on scales we are
interested in (several comoving Mpc) are not deterministic functions of the
gas properties, and must be treated statistically.

In addition, since our box volumes are still relatively small and most quasars
found in our simulation boxes are still too faint to be detected by even the
best telescopes, constrained simulations can be used to model \HII\ regions
around the quasars comparable in luminosity to those observed by the SDSS.

This weak dependence of the simulation results on the particular choice of clumping factors is the main justification of our approach. It demonstrates that, as long as ionization and recombinations are counted self-consistently, the specific form of spatial averaging used to define clumping factors is not too important.

\acknowledgements
This work was largely motivated by Jordi Miralda-Escud\'{e} and we are grateful to him for numerous discussions. We also thank Xiaohui Fan for providing data, offering information and stimulating the discussion. The project was supported by the NASA astrophysical theory program grant NAG5-12029 and we used the SGI Origin 2000 array and the IBM P690 array at the National Center for Supercomputing Applications. This work was supported by the DOE and the NASA grant NAG 5-10842 at Fermilab.

\end{document}